%% file: main.tex
\documentclass[longauth]{aa}

\usepackage[colorlinks=true, linkcolor=blue, citecolor=blue, filecolor=blue, urlcolor=blue]{hyperref}

\usepackage{txfonts}

\usepackage{graphicx}
\usepackage{natbib}
\bibpunct{(}{)}{;}{a}{}{,} 

\usepackage{lscape}

\usepackage[colorlinks=true, linkcolor=blue, citecolor=blue, filecolor=blue, urlcolor=blue]{hyperref}

\newcommand\textlcsc[1]{\textsc{\MakeLowercase{#1}}}

\newcommand{\lya}{Ly$\alpha$\xspace}
\newcommand{\angstrom}{\textup{\AA}\xspace}
\newcommand{\rew}{REW$_{\rm Ly\alpha}$\xspace}

\newcommand{\refdel}{\color{black}}

\begin{document}

\title{JADES: The emergence and evolution of Ly$\alpha$ emission \& constraints on the IGM neutral fraction}

\author{
Gareth C. Jones\inst{1}\thanks{E-mail: gareth.jones@physics.ox.ac.uk}
\and
Andrew J. Bunker\inst{1}\and
Aayush Saxena\inst{1,2}\and
Joris Witstok\inst{3,4}\and
Daniel P. Stark\inst{5}\and
Santiago Arribas\inst{6}\and
William M. Baker\inst{3,4}\and
Rachana Bhatawdekar\inst{7,8}\and
Rebecca Bowler\inst{9}\and
Kristan Boyett\inst{10,11}\and
Alex J. Cameron\inst{1}\and
Stefano Carniani\inst{12}\and
Stephane Charlot\inst{13}\and
Jacopo Chevallard\inst{1}\and
Mirko Curti\inst{14,3,4}\and
Emma Curtis-Lake\inst{15}\and
Daniel J. Eisenstein\inst{16}\and
Kevin Hainline\inst{5}\and
Ryan Hausen\inst{17}\and
Zhiyuan Ji\inst{5}\and
Benjamin D. Johnson\inst{16}\and
Nimisha Kumari\inst{18}\and
Tobias J. Looser\inst{3,4}\and
Roberto Maiolino\inst{3,4,2}\and
Michael V. Maseda\inst{19}\and
Eleonora Parlanti\inst{12}\and
Hans-Walter Rix\inst{20}\and
Brant E. Robertson\inst{21}\and
Lester Sandles\inst{3,4}\and
Jan Scholtz\inst{3,4}\and
Renske Smit\inst{22}\and
Sandro Tacchella\inst{3,4}\and
Hannah \"{U}bler\inst{3,4}\and
Christina C. Williams\inst{23}\and
Chris Willott\inst{24}
}

\institute{
Department of Physics, University of Oxford, Denys Wilkinson Building, Keble Road, Oxford OX1 3RH, UK\label{1}
\and
Department of Physics and Astronomy, University College London, Gower Street, London WC1E 6BT, UK\label{2}
\and
Kavli Institute for Cosmology, University of Cambridge, Madingley Road, Cambridge CB3 0HA, UK\label{3}
\and
Cavendish Laboratory, University of Cambridge, 19 JJ Thomson Avenue, Cambridge CB3 0HE, UK\label{4}
\and
Steward Observatory, University of Arizona, 933 N. Cherry Ave., Tucson, AZ 85721 USA\label{5}
\and
Centro de Astrobiolog\'{i}a (CAB), CSIC–INTA, Cra. de Ajalvir Km.~4, 28850- Torrej\'{o}n de Ardoz, Madrid, Spain\label{6}
\and
European Space Agency (ESA), European Space Astronomy Centre (ESAC), Camino Bajo del Castillo s/n, 28692 Villanueva de la Ca\~{n}ada, Madrid, Spain\label{7}
\and
European Space Agency, ESA/ESTEC, Keplerlaan 1, 2201 AZ Noordwijk, NL\label{8}
\and
Jodrell Bank Centre for Astrophysics, Department of Physics and Astronomy, School of Natural Sciences, The University of Manchester, Manchester, M13 9PL, UK\label{9}
\and
School of Physics, University of Melbourne, Parkville 3010, VIC, Australia\label{10}
\and
ARC Centre of Excellence for All Sky Astrophysics in 3 Dimensions (ASTRO 3D), Australia\label{11}
\and
Scuola Normale Superiore, Piazza dei Cavalieri 7, I-56126 Pisa, Italy\label{12}
\and
Sorbonne Universit\'{e}, CNRS, UMR 7095, Institut d'Astrophysique de Paris, 98 bis bd Arago, 75014 Paris, France\label{13}
\and
European Southern Observatory, Karl-Schwarzschild-Strasse 2, 85748 Garching, Germany\label{14}
\and
Centre for Astrophysics Research, Department of Physics, Astronomy and Mathematics, University of Hertfordshire, Hatfield AL10 9AB, UK\label{15}
\and
Center for Astrophysics $|$ Harvard \& Smithsonian, 60 Garden St., Cambridge, MA 02138, USA\label{16}
\and
Department of Physics and Astronomy, The Johns Hopkins University, 3400 N. Charles St., Baltimore, MD 21218, USA\label{17}
\and
AURA for European Space Agency, Space Telescope Science Institute, 3700 San Martin Drive. Baltimore, MD, 21210, USA\label{18}
\and
Department of Astronomy, University of Wisconsin-Madison, 475 N. Charter St., Madison, WI 53706 USA\label{19}
\and
Max-Planck-Institut f\"ur Astronomie, K\"onigstuhl 17, D-69117, Heidelberg, Germany\label{20}
\and
Department of Astronomy and Astrophysics, University of California, Santa Cruz, 1156 High Street, Santa Cruz, CA 95064, USA\label{21}
\and
Astrophysics Research Institute, Liverpool John Moores University, 146 Brownlow Hill, Liverpool L3 5RF, UK\label{22}
\and
NSF’s National Optical-Infrared Astronomy Research Laboratory, 950 North Cherry Avenue, Tucson, AZ 85719, USA\label{23}
\and
NRC Herzberg, 5071 West Saanich Rd, Victoria, BC V9E 2E7, Canada\label{24}
}

\date{Received X / Accepted Y}

\abstract
{The rest-frame UV recombination emission line \lya can be powered by {\refdel ionising} photons from young massive stars in star forming galaxies, but its ability to be resonantly scattered by neutral gas complicates its interpretation. For reionization era galaxies, a neutral intergalactic medium (IGM) will scatter \lya from the line of sight, making \lya a useful probe of the neutral fraction evolution. Here, we explore \lya in JWST/NIRSpec spectra from the ongoing JADES {\refdel programme}, which targets hundreds of galaxies in the well-studied GOODS-S and GOODS-N fields. These sources are UV-faint ($-20.4<\rm M_{\rm UV}<-16.4$), and thus represent a poorly-explored class of galaxies. The low spectral resolution ($R\sim100$) spectra of a subset of 84 galaxies in GOODS-S with $z_{spec}>5.6$ (as derived with optical lines) are fit with line and continuum models, in order to search for significant line emission. Through exploration of the R100 data, we find evidence for \lya in 17 sources. This sample allows us to place observational constraints on the fraction of galaxies with \lya emission in the redshift range $5.6<z<7.5$, with a decrease from $z=6$ to $z=7$. We also find a positive correlation between \lya equivalent width and M$_{UV}$, as seen in other samples. These results are used to estimate the neutral gas fraction at $z\sim7$, agreeing with previous results ($X_{HI}\sim0.5-0.9$). }

\keywords{(cosmology:) dark ages, reionization, first stars - (galaxies:) intergalactic medium - galaxies: high-redshift}
\maketitle

\section{Introduction}\label{intro}

By studying the properties of galaxies at high redshift (such as morphology, spectral energy distributions, and kinematics), we are able to chart how populations of galaxies have evolved through cosmic time. In individual galaxies we can study the buildup of gaseous reservoirs, the conversion of this fuel into stars, and the effects of feedback. By studying the overall galaxy population as a function of redshift, we can determine the evolution of the luminosity function, the star formation rate density, and the growth of supermassive black holes. In parallel, these studies shine light on the last great phase transition of the Universe, when the intergalactic medium (IGM) became {\refdel ionised} (i.e., the epoch of reionization; EoR).

This epoch began at the end of the `cosmic dark ages', when the first stars formed (e.g., \citealt{vill18}). The UV radiation of these objects created {\refdel ionised} regions (i.e., `bubbles'), which grew and merged together (e.g., \citealt{gned00}). Observations suggest that the IGM was mostly {\refdel ionised} at $z\sim6$ ($t_H\sim0.91$\,Gyr; e.g., \citealt{fan06}), although the details of reionization are still being derived (e.g., the drivers; \citealt{hutc19,naid20,ends21}, topology; \citealt{pent14,lars22,yosh22}, and timeline; \citealt{chri21,cain21,zhu22}). One of the most useful tools for studying this epoch is the bright Lyman-$\alpha$ line of hydrogen ($\lambda=1215.67\angstrom$; hereafter \lya).

As the lowest-energy transition ($n=2\rightarrow1$) of the most abundant element, \lya emission should be ubiquitous. But this radiation may be absorbed and re-radiated by any other hydrogen atom in the ground state (i.e., HI). For galaxies at $z\lesssim6$, this repeated absorption and re-radiation by neutral gas inside a galaxy (i.e., resonant scattering) means that \lya can be greatly reduced in intensity, but also may be observed along sight lines distant from the original emission region, as seen in large \lya halos of $\sim10$\,kpc (e.g., \citealt{drak22,kiku23}) or $\sim100$\,kpc (e.g., \citealt{stei00,reul03,dey09,cai17,li21,guo23,zhan23}). 

For galaxies in the EoR, neutral gas in the IGM surrounding a galaxy may also scatter \lya emission, resulting in a lower observed brightness (e.g., \citealt{font10,star10}). In order for this emission to be observable, it must lie in an {\refdel ionised} bubble (e.g., \citealt{maso20}) and/or feature a significant outflow (e.g., \citealt{dijk10}). So by comparing the fraction of galaxies with \lya emission to the expected number from models (\lya fraction; $X_{Ly\alpha}$), we are able to place constraints on the HI filling fraction (X$_{\rm HI}$; e.g., \citealt{ono12,maso18,matt22}). 

Constraints on $X_{Ly\alpha}$ have been placed for galaxies from $4\lesssim z\lesssim8$ (e.g., \citealt{star11,curt12,caru12,caru14,ono12,sche14,star17,pent18,yosh22}) and down to $z\sim2$ (e.g., \citealt{cass15}). By comparing the observed evolution of $X_{Ly\alpha}$ to that expected from different \lya luminosity functions, some of these works have placed constraints on X$_{\rm HI}$, suggesting that it quickly decreased from $\gtrsim0.9$ to $\sim0$ between $z\sim8$ and $z\sim6$ (e.g., \citealt{maso18,maso19,mora21}). Therefore, the study of galaxies in this short time interval ($\sim0.3$\,Gyr) is key to {\refdel characterising} the timeline of reionization. While a number of studies have been undertaken, observations have been hampered by small sample size, limited volume volume (prone to cosmic variance), or a focus on bright ($M_{UV}\lesssim-20$) or strongly lensed sources (e.g., \citealt{hoag19,full20,bola22}). Already, {\refdel the JWST/Near-Infrared Spectrograph (NIRSpec; \citealt{jako22,boke23})} has seen great success in detecting \lya (e.g., \citealt{bunk23a,jung23,roy23,tang23}). But to reduce sample variance and allow stronger conclusions, a wide-area survey down to $M_{UV}\sim-18.75$ is needed (e.g., \citealt{tayl14}). With JWST, this survey is now possible. 

The JWST Advance Deep Extragalactic Survey (JADES; \citealt{bunk20,eise23}) is a cycle 1-2 GTO {\refdel programme} observing the GOODS (The Great Observatories Origins Deep Survey; \citealt{dick03}) north and south fields with {\refdel JWST/NIRspec} in multi-object spectroscopy mode (\citealt{ferr22}) {\refdel in both low spectral resolution (R100) and medium spectral resolution (R1000)}, in combination with JWST/Near-Infrared Camera (NIRCam; \citealt{riek23}). 

This rich dataset is the subject of numerous ongoing investigations, including detailed {\refdel modelling} of the \lya profiles using the R1000 spectra (e.g., asymmetry, velocity offsets; \citealt{saxe23}), analysis of the damping wings (Jakobsen et al. in prep), and a search for \lya overdensities that hint at large {\refdel ionised} bubbles (\citealt{wits23}). In this work, we search for \lya emission in the R100 spectra of data from GOODS-S for the purpose of placing constraints on the neutral gas fraction at $z\sim6-8$. These fits are used to examine correlations between \lya rest-frame equivalent width (\rew), redshift, and UV absolute magnitude. 

We describe our sample in Section \ref{sec_sample}. The details and results of our R100 spectral fitting procedure are given in Section \ref{sec_fit}. These findings are discussed in Section \ref{sec_disc}, and we conclude in Section \ref{sec_conc}. We assume a standard concordance cosmology ($\Omega_{\Lambda}$,$\Omega_m$,h)=(0.7,0.3,0.7) throughout.

\section{Sample}\label{sec_sample}
\subsection{Observations overview}
JADES consists of two survey depths (i.e., `Deep' and `Medium'). The former allows for {\refdel characterisation} of a small number of dimmer galaxies or more detailed study of individual sources at higher S/N, while the latter enables a statistical {\refdel characterisation} of the galaxy population at high-\textit{z}. In addition, each tier has two stages with different selections: one based on existing {\refdel Hubble Space Telescope (}HST{\refdel )} imaging (followed by `/HST') and the other based on JWST/NIRCam imaging (followed by `/JWST'; see \citealt{eise23} for more detail).

From the JADES survey, we {\refdel utilised} data from galaxies in GOODS-S in the Deep/HST (PID: 1210, PI: N. L\"{u}tzgendorf), Medium/HST (PID: 1180, PI: D. Eisenstein), and Medium/JWST (PID: 1286, PI: N. L\"{u}tzgendorf) subsurveys. Target {\refdel catalogues} were created for each tier, with galaxies assigned priority classes (PCs) based on photometric redshift, apparent UV-brightness, and visual inspection of existing ancillary data (for more details on Deep/HST priority classes, see \citealt{bunk23b}). Galaxies with $z_{phot}>5.6$ were collected from studies that selected sources based on the Lyman-break drop out selection (e.g., \citealt{bunk04,bouw15,hari16}), or Lyman-break galaxies (LBGs). This system ensures that both rare (e.g., bright, high-redshift, {\refdel hosts of active galactic nuclei}) and representative systems would be observed. Due to the geometrical constraints dictated by mask construction, galaxies were randomly selected for observation from each PC.

Details of the data acquisition and reduction details are given in other works (\citealt{curt23}, Carniani et al. in prep), which we {\refdel summarise} here. Targets were observed with three shutters in a three-point nod. In order to improve data quality, sub-pointings were created for each primary pointing by shifting the {\refdel JWST/NIRSpec Multi-Shutter Array (}MSA{\refdel )} by a few shutters in each direction. Due to failed shutters, not all targets were observable in all three sub-pointings. This results in exposure times of sources in R100 for Deep/HST of 33.6-100.8\,ks for 253\,observed sources. Medium/JWST features a similar setup, but with a lower R100 exposure time per target: 5.3-8.0\,ks for 169\,observed sources. While 1354\,galaxies were observed in Medium/HST, the majority of the executions (8/12) were negatively affected by a short circuit in the MSA, making the data unusable. The mean exposure time for R100 per object of the usable data is $3.8$\,ks (\citealt{eise23}). Observations were repeated for some of these objects with unusable data (364\,sources), with a mean R100 exposure time per object of $7.5-11.3$\,ks.

The resulting raw data were calibrated using a pipeline developed by the ESA NIRSpec Science Operations Team (SOT) and the NIRSpec GTO Team, which includes corrections for outlier rejection (e.g., `snowballs'), background subtraction (using adjacent slits), wavelength grid resampling, and slit loss. This results in 2D spectra with data quality flags, which was used to extract a 1D spectrum and a noise spectrum.

\subsection{Sample construction}
\noindent
The resulting spectra of all observed sources were visually inspected and strong emission lines (e.g., [OIII]$\lambda$5007, H$\alpha$) were fit. We impose a lower redshift limit of $z_{spec}>5.6$ in order to ensure a sample of LBGs. This yields a sample of 84 galaxies at $z\gtrsim5.6$ (38 from Medium/HST, 13 from Medium/JWST, and 33 from Deep/HST) with precise spectroscopic redshifts (full details in \citealt{bunk23b}). Both R100 and R1000 spectra are available for these galaxies, and we proceed with the R100 spectra in this work\footnote{Details of the R1000 analysis are presented in an associated paper (\citealt{saxe23})}.

Our sample of 84\,galaxies is composed of galaxies at $z_{spec}>5.6$ (i.e., LBGs) with cuts on UV brightness in HST broadband filters redward of the Lyman break. Some sources were excluded from our sample for a lack of strong emission lines (i.e., a poorly {\refdel constrained} $z_{spec}$). While this results in a more complex sample than the uniform sample selection of some previous studies (e.g., \citealt{pent18,yosh22}), this inhomogeneity is taken into account through the error spectrum of each source and a completeness analysis.

\section{Combined Lya and continuum fit}\label{sec_fit}
At these high redshifts, the spectra exhibit a strong continuum break at the \lya wavelength at the redshift of the galaxy. The deep sensitivity of the R100 data allows us to simultaneously {\refdel characterise} any Ly$\alpha$ emission and the underlying continuum of each source. To do this, we examine whether a two-component model (i.e., line and continuum) or a single-component model (i.e., only continuum) better fits the extracted spectrum of each source using \textlcsc{lmfit} \citep{newv14}. This process is detailed below.

\subsection{Resolution effects}\label{reseff}

Due to the low spectral resolution of the R100 data, we must consider both the wavelength grid and spectral dispersion. The spectral pixels in our calibrated data are large ($\Delta$v$\sim 2000-2600$\,km\,s$^{-1}$ per pixel at the redshifted \lya wavelength for galaxies at $z\sim5.6-10$). Furthermore, for galaxies observed in the EoR this wavelength is near the minimum of the PRISM resolving power curve\footnote{As recorded in the JWST documentation; \url{https://jwst-docs.stsci.edu/jwst-near-infrared-spectrograph/nirspec-instrumentation/nirspec-dispersers-and-filters}}, with $R\sim30$. This implies that the line-spread function (LSF) has a full width at half maximum (FWHM) of $\sim10^4$\,km\,s$^{-1}$. Although for a compact source which does not fill the slit, the resolving power will in practice be higher by as much as a factor of 2 (De Graaff et al. in prep).

The low resolution also makes it impossible to {\refdel characterise} the \lya profile (e.g., asymmetry, velocity offset). Instead, the \lya emission may be approximated as additional flux in the first spectral bin redward of the \lya break, which is spread into {\refdel neighbouring} bins by the LSF. 

\begin{figure}
\centering
\includegraphics[width=0.5\textwidth]{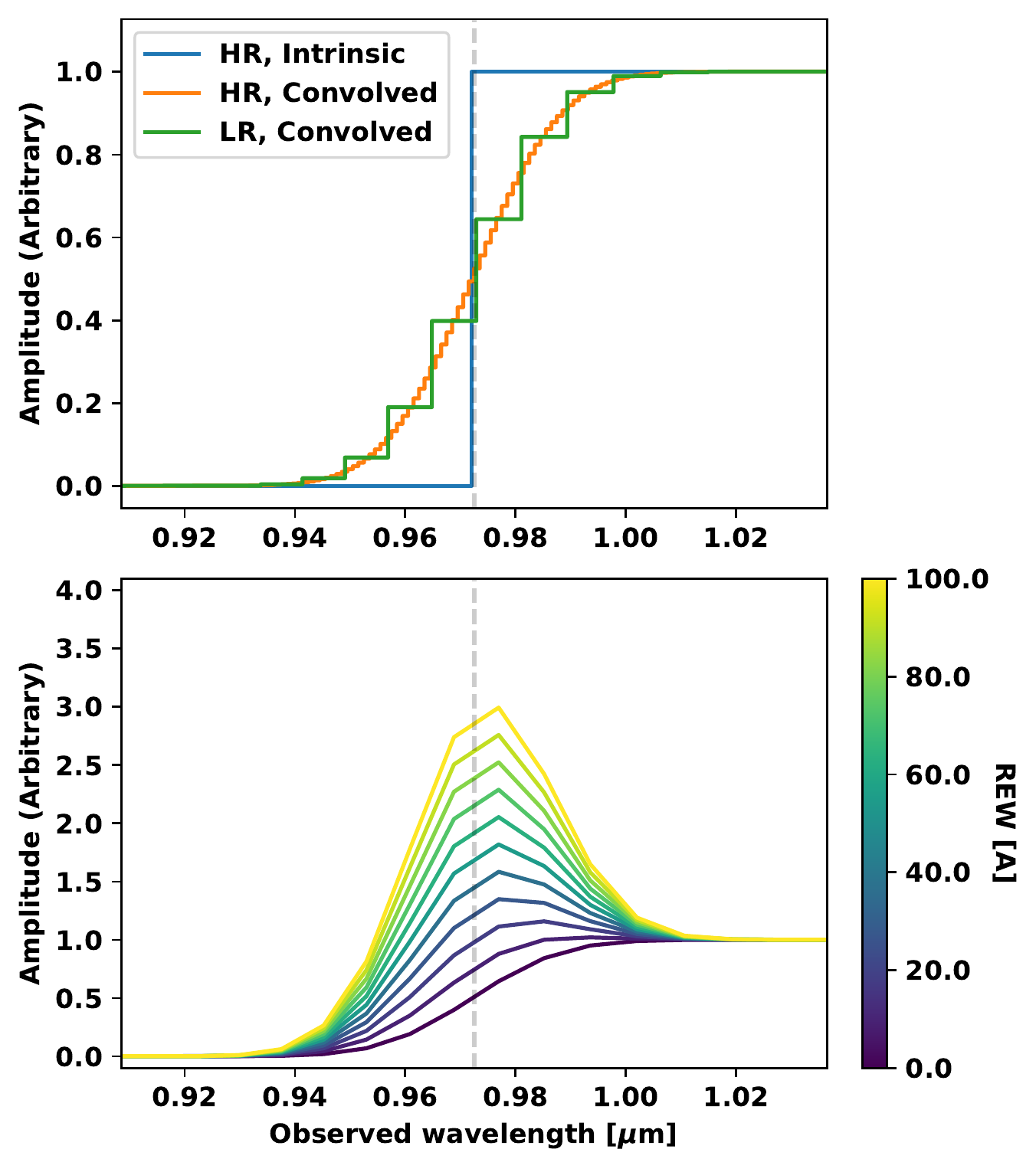}
\caption{Demonstration of how \lya break and emission for a source at $z=7$ are affected by the low resolving power of our observations. The top panel shows how a step function (blue line) is affected by the resolving power  on a high-resolution ($\Delta \lambda=0.001\,\mu$m) spectral grid (orange curve), and how this curve would appear on the R100 spectral grid (green steps). If we add \lya emission with a given \rew in the first high-resolution spectral bin redwards of the break and then account for the LSF and R100 spectral bin, we find the curves in the lower panel.}
\label{conv_ex}
\end{figure}

To demonstrate how this affects the interpretation of R100 spectra, we first create a higher-resolution ($\Delta \lambda=0.001$\,$\mu$m, or $R\sim730$) model of a \lya break ({\refdel modelled} as a step function) at $z=7$ with no \lya flux (blue line in the top panel of Figure \ref{conv_ex}). If we account for the LSF by convolving the spectrum with a Gaussian ($\sigma_R=\lambda_{Ly\alpha}/R/2.355$), the break becomes an S-shaped curve instead (orange histogram). Rebinning this curve to the coarser R100 wavelength grid maintains the curve, but at lower resolution (green histogram).

If we add \lya flux with a given \rew to the intrinsic high-resolution model as additional flux in the first spectral bin redward of the \lya break, convolve the model with the LSF, and rebin the result to the R100 spectral grid, we find the profiles shown in the lower panel of Figure \ref{conv_ex}. The line flux is spread from one low-resolution pixel into a Gaussian that spans both sides of the \lya break. Even high-\rew lines (e.g., 100\,$\angstrom$) have low peaks (here $4\times$ the continuum level). In addition, low-\rew lines ($<20\angstrom$) feature very low amplitudes, and instead appear similar to a pure continuum model with a blueshifted \lya break.

With this in mind, our model fitting procedure begins with a high-resolution spectral model, which is convolved with a Gaussian to account for the resolving power and then rebinned to the R100 spectral grid. This allows us to compare the observed and model spectra directly, in order to extract the intrinsic continuum and \lya flux.

\subsection{Model description}

We first assume that the underlying continuum can be approximated by a power law and use a Heaviside step function to represent the \lya break (see Appendix \ref{dampapp} for a discussion of this assumption). This continuum-only model only features two variables: the continuum value at a rest-frame wavelength of $1500\angstrom$ ($S_{C,o}$) and the spectral slope just redwards of \lya ($n$; $\sim1300-1500\,\angstrom$ rest-frame), which is not fixed to the redder (i.e., $\sim1400-2300\,\angstrom$ rest-frame) spectral slope $\beta$ derived by \citet{saxe23}.

In the case where both continuum and \lya emission are detected, the line emission will have a rest-frame equivalent width:
\begin{equation}\label{ew0}
REW_{Ly\alpha}=\frac{F_{Ly\alpha}}{(1+z)S_{C}(\lambda_{Ly\alpha,obs})}
\end{equation}
where $F_{Ly\alpha}$ is the total line flux of \lya. As discussed in Section \ref{reseff}, the low spectral resolution of the R100 data dictates that our line emission model is simple. Our combined line and continuum model thus has three variables: those of the continuum model (i.e., $S_{C,o}$ and $n$) and \rew.

For both the continuum and line+continuum models, we first create a spectral grid of high resolution ($\Delta \lambda=0.001\,\mu$m) and populate each bin using a continuum-only or continuum and line model. As discussed in Section \ref{reseff}, we then convolve the spectrum with a Gaussian that accounts for the LSF. We first consider using a Gaussian based on the theoretical resolving power as recorded in the JWST documentation ($\sigma_R=\lambda_{Ly\alpha}/R/2.355$). However, this was calculated assuming a source that illuminates the slit uniformly, which is not the case for the relatively compact sources in our sample. Detailed LSFs for the sources in Deep/HST have been calculated (de Graaff et al. in prep), which reveal that the actual LSF is smaller than the theoretical value, by a factor of up to $\sim2.4$. However these models are not available for our whole sample. To account for the LSF in a uniform manner, we convolve the model spectrum with a Gaussian of width $F_R\sigma_R$, where $F_R$ is allowed to vary. 
These data are then rebinned to the R100 spectral grid of an observation.

\subsection{Fitting procedure}\label{fitsec}

We use \textlcsc{lmfit} with a `leastsq' {\refdel minimiser} to fit each convolved model to a subset of the observed spectrum
Each data point is weighted by its associated inverse variance (as derived from the error spectrum).

We limit the fit subset to the wavelength range $\left[\left(\lambda_{Ly\alpha}\pm(0.03\mu\mathrm{m})\right)\times(1+z)\right]$, with a minimum of $\ge0.75\,\mu$m. This range is chosen to avoid including excessive amounts of noisy data at blue wavelengths below the \lya break, and to only fit the continuum just redwards of \lya, avoiding nearby emission lines (e.g., [CIV]$\lambda\lambda1548,1551$, HeII$\lambda1640$).

While precise systemic redshifts for each source have been derived using fits to strong lines (e.g., [OIII]$\lambda5007$, H$\alpha$) using the higher spectral resolution gratings (\citealt{bunk23b}), it is possible that the \lya emission is shifted into a {\refdel neighbouring} spectral bin by a large velocity offset (i.e., up to a few hundred km\,s$^{-1}$; \citealt{erb14,marc14}) or a different binning scheme between the gratings and prism. This is accounted for by allowing the redshift of \lya emission to vary from the systemic redshift within the R100 bin, taking the result with the lowest $\chi^2$. 

Next, we consider the lowest REW line that we can detect for each source. Because the line emission is spread from one into multiple channels by convolution with the LSF, we may approximate the $1\sigma$ limit on \rew in the R100 spectrum as:
\begin{equation}\label{drew}
\Delta REW_{Ly\alpha,1\sigma}=\frac{\sqrt{2\pi}E(\lambda_{Ly\alpha,obs})F_R\sigma_R}{(1+z)S_{C,Ly\alpha}}
\end{equation}
where $F_R\sigma_R$ is the width of the LSF and $E(\lambda)$ is the error spectrum.

The results of the continuum (`C') and line and continuum (`L+C') are examined, and a definite \lya detection is reported if both of the following criteria are met:
\begin{itemize}
    \item The `L+C' fit features a lower $\chi_{red}^2$ than the `C' fit.
    \item The best-fit \rew is greater than $2.5\Delta REW_{Ly\alpha,1\sigma}$.
\end{itemize}
When \lya is detected, we take the best-fit \rew value and its associated uncertainty from our fit. Otherwise, we treat the \lya line as undetected, and use $3\Delta REW_{Ly\alpha,1\sigma}$ as an upper limit. 

\subsection{Results}
\begin{figure*}
\centering
\includegraphics[width=0.49\textwidth]{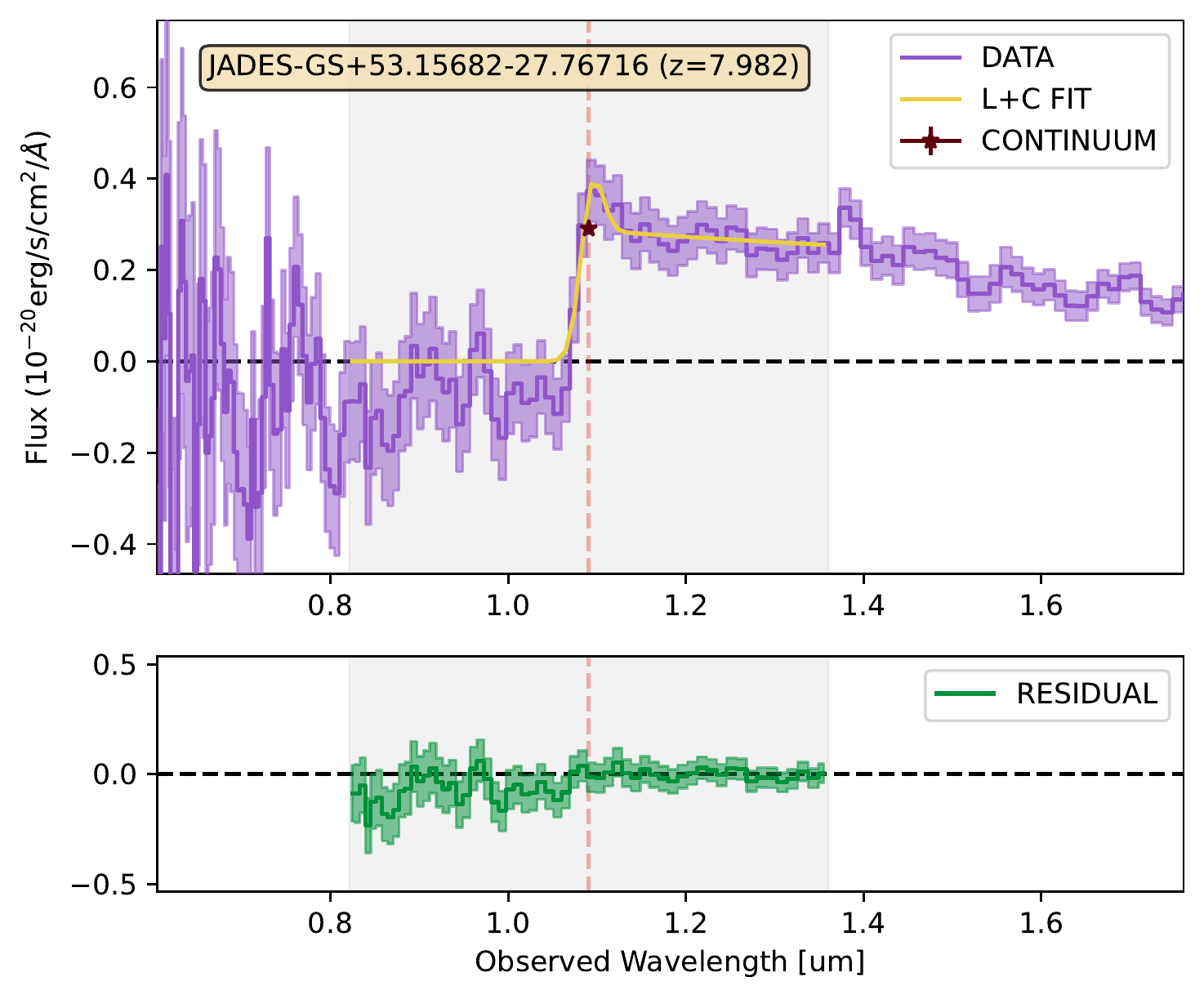}
\includegraphics[width=0.49\textwidth]{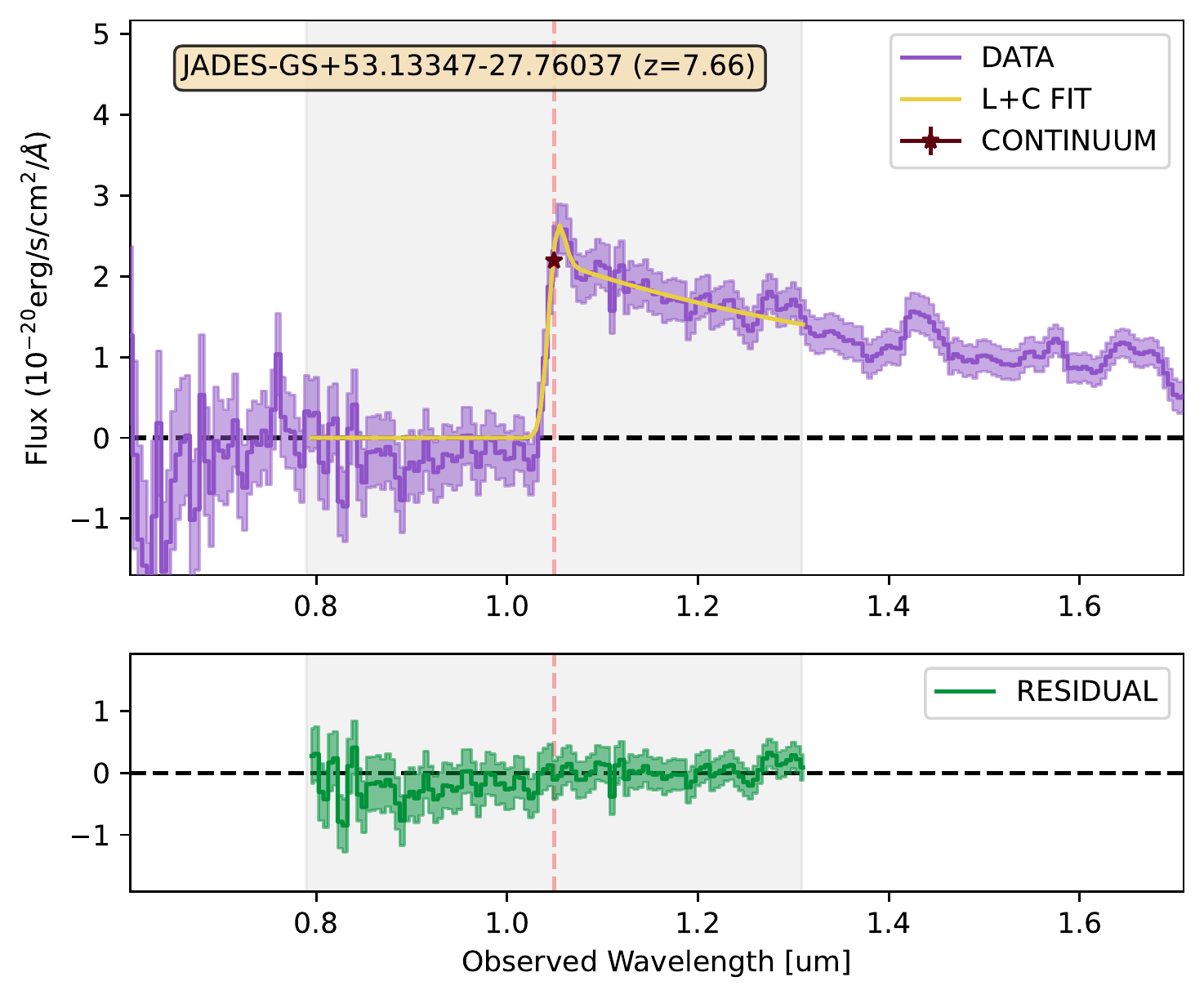}
\includegraphics[width=0.49\textwidth]{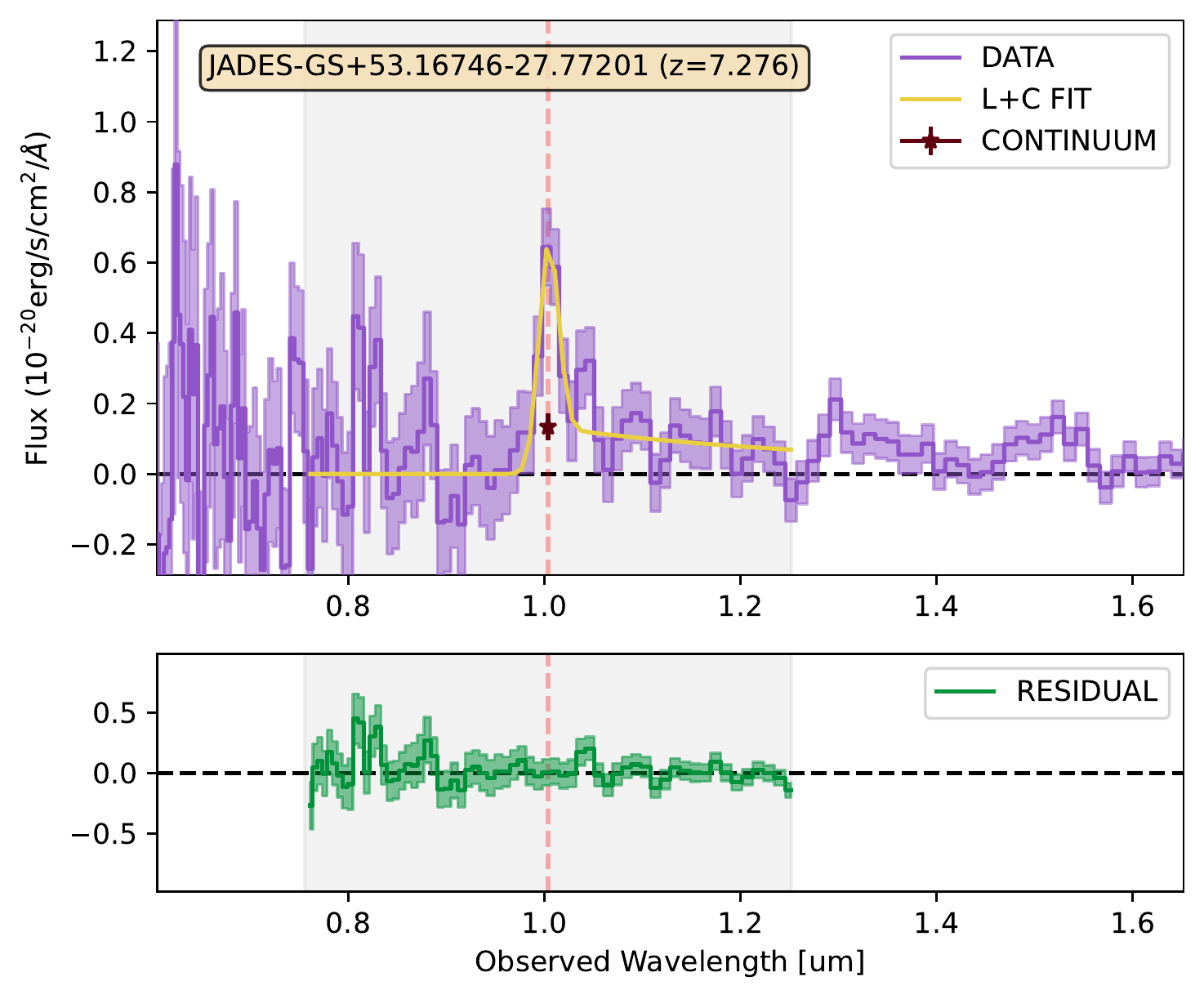}
\includegraphics[width=0.49\textwidth]{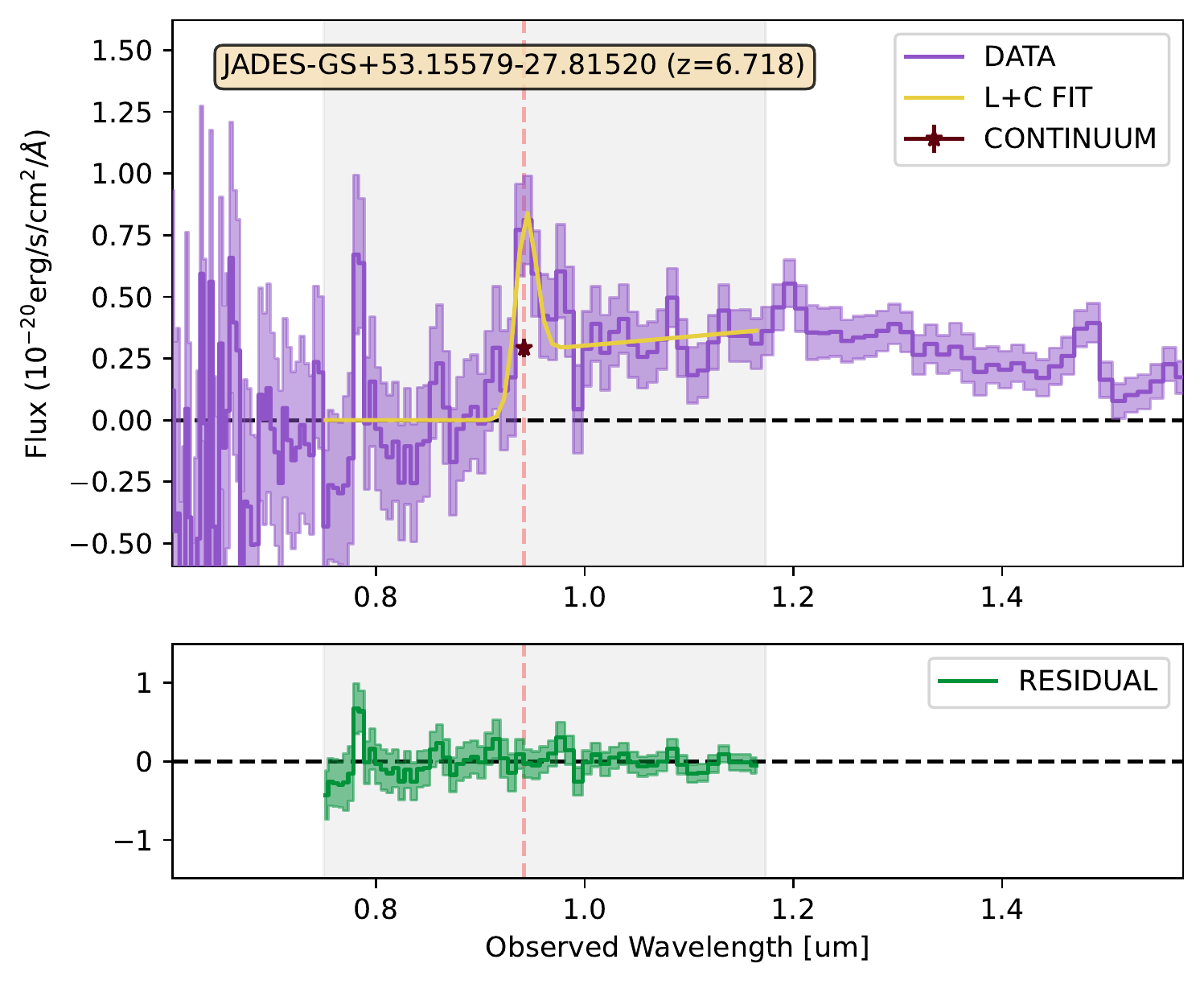}
\includegraphics[width=0.49\textwidth]{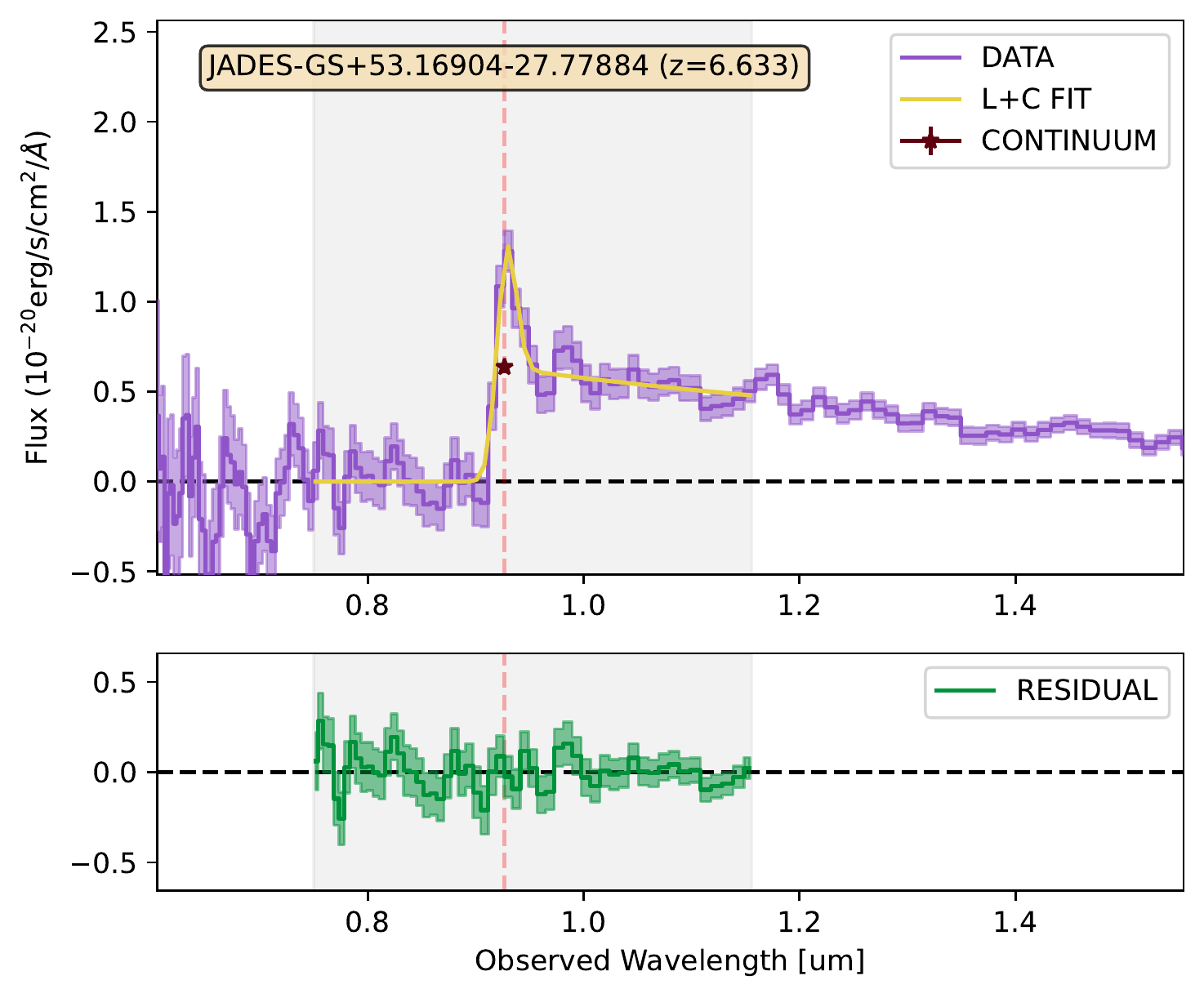}
\includegraphics[width=0.49\textwidth]{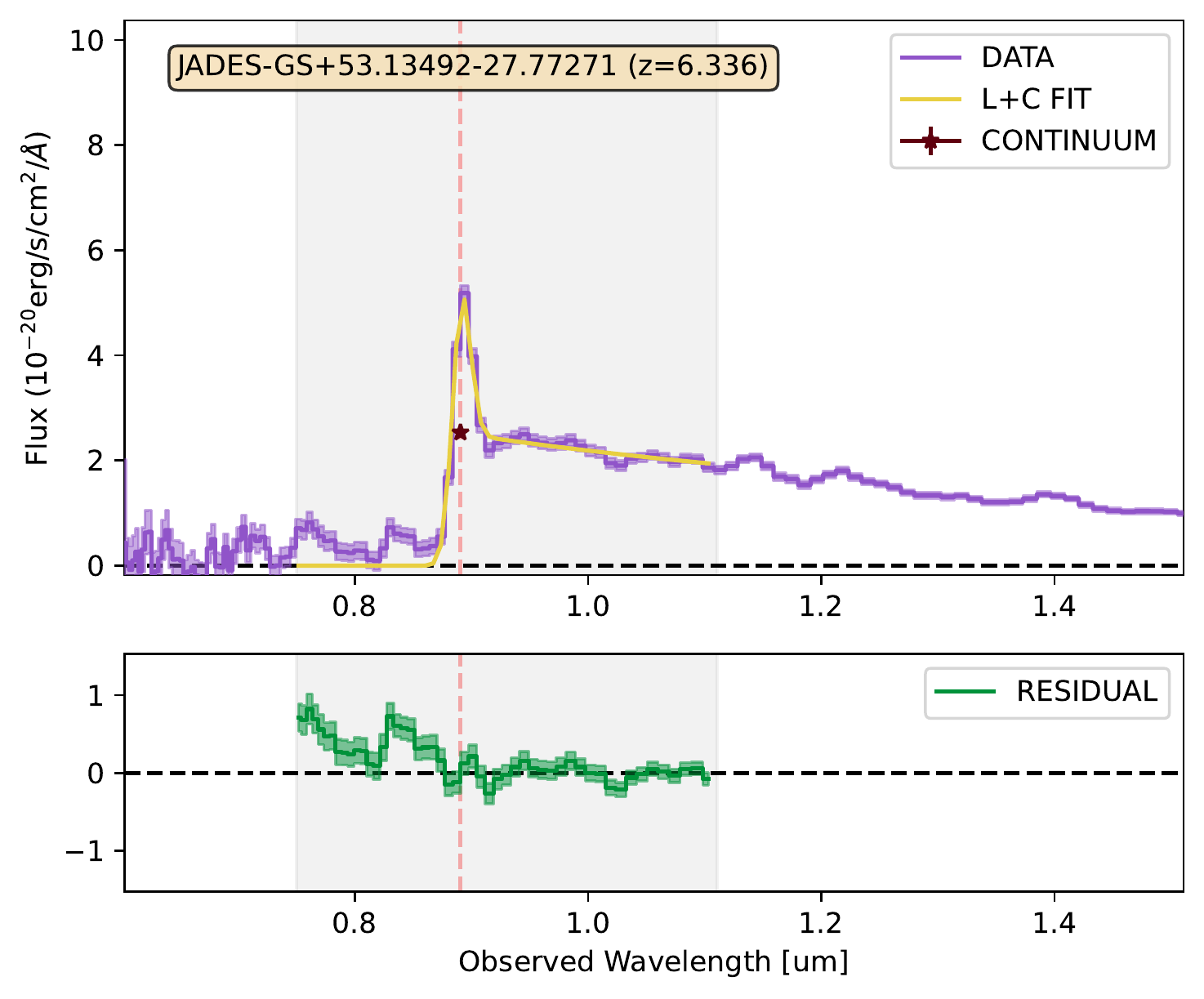}

\caption{Results of fitting a line + continuum model to observed JADES R100 data, for sources detected in \lya emission (denoted by `L+C FIT'). 
In each top panel, we show the observed spectrum (purple line) with an associated $1\sigma$ error (shaded region). The best-fit model, which includes the effects of the LSF, is shown by a yellow line. Fitting was performed using the wavelength range that is shaded grey. The continuum value at the redshifted \lya wavelength is represented by a brown star. The bottom panel shows the residual.}
\label{lyares1}
\end{figure*}

\begin{figure*}
\centering
\includegraphics[width=0.49\textwidth]{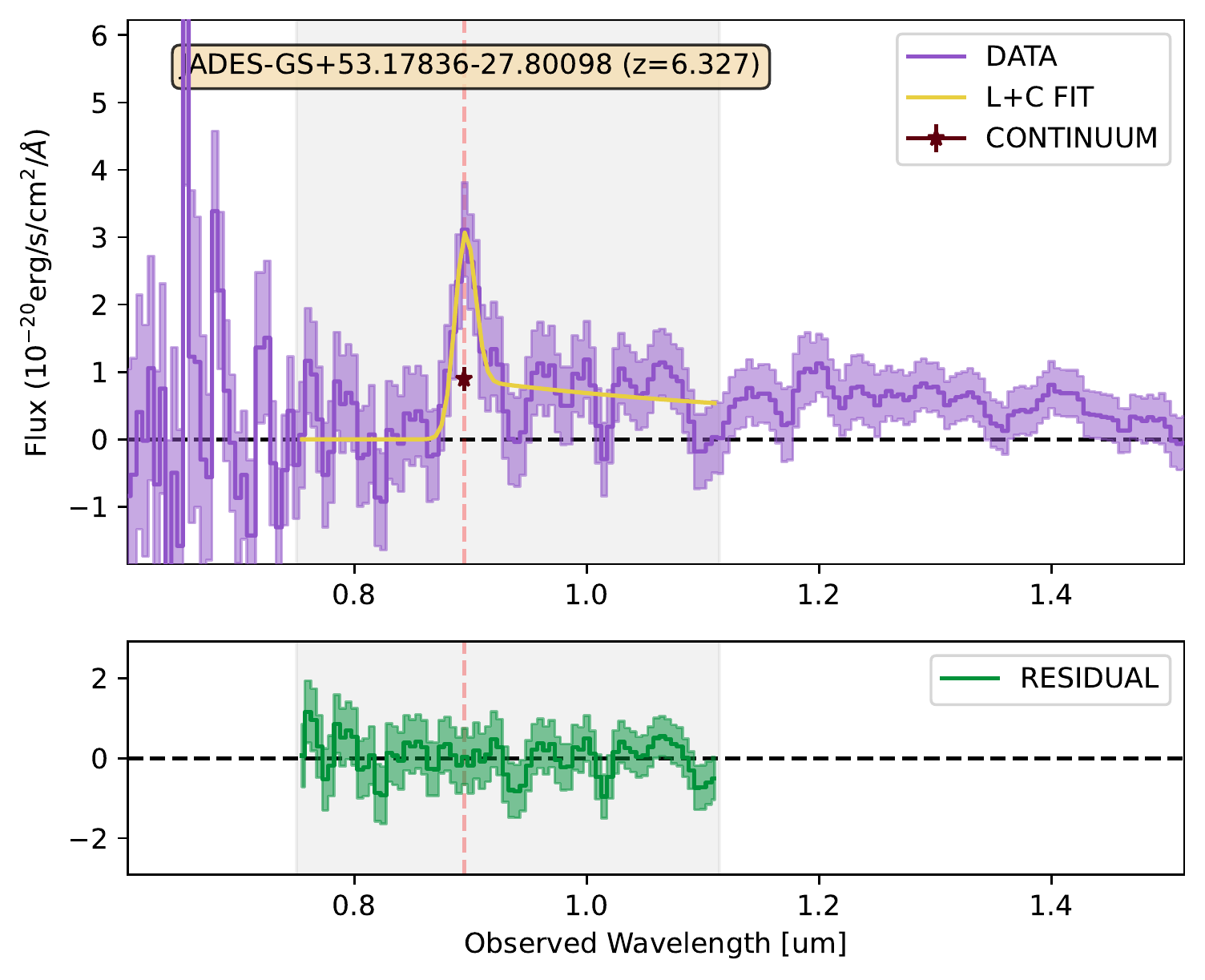}
\includegraphics[width=0.49\textwidth]{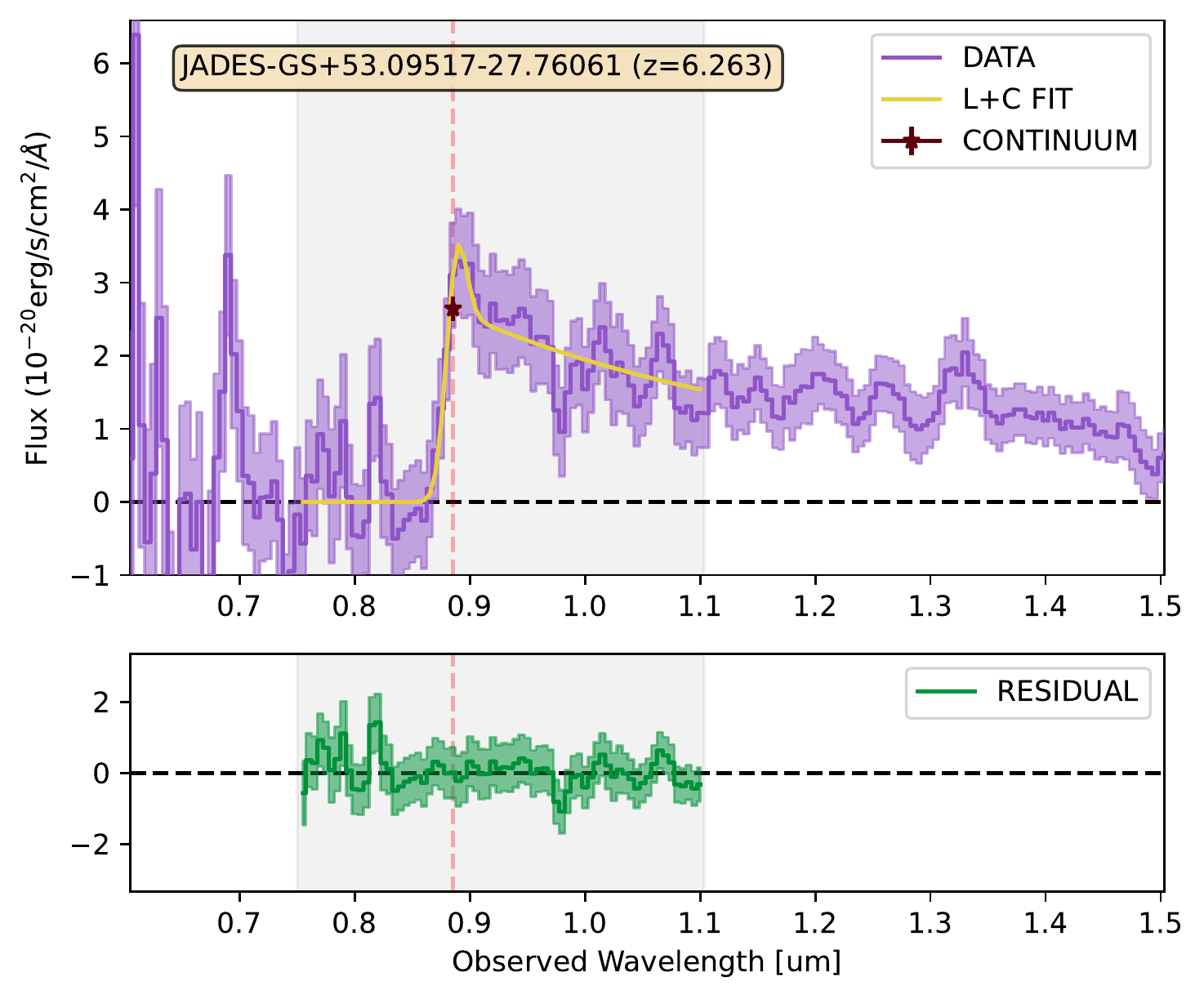}
\includegraphics[width=0.49\textwidth]{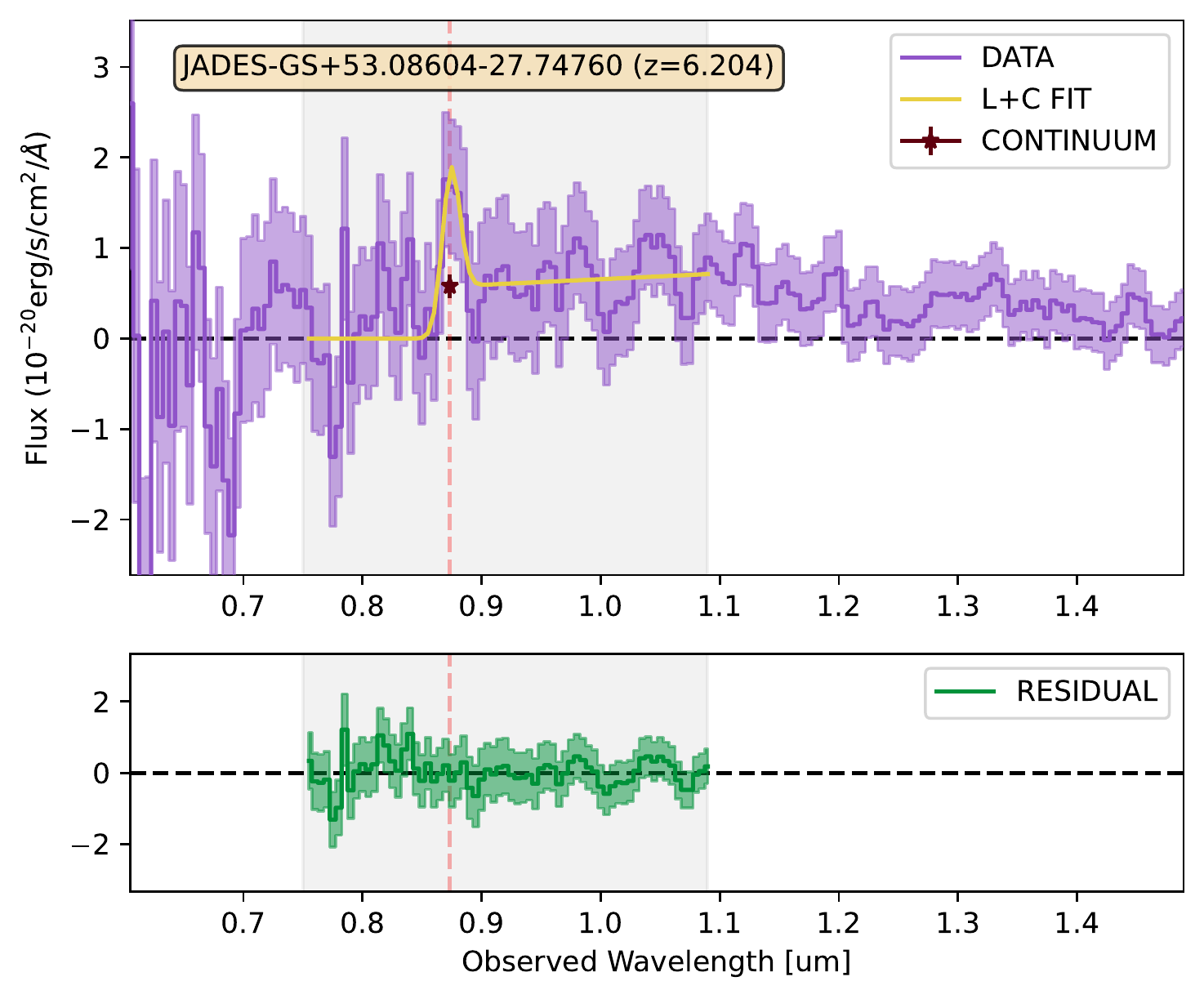}
\includegraphics[width=0.49\textwidth]{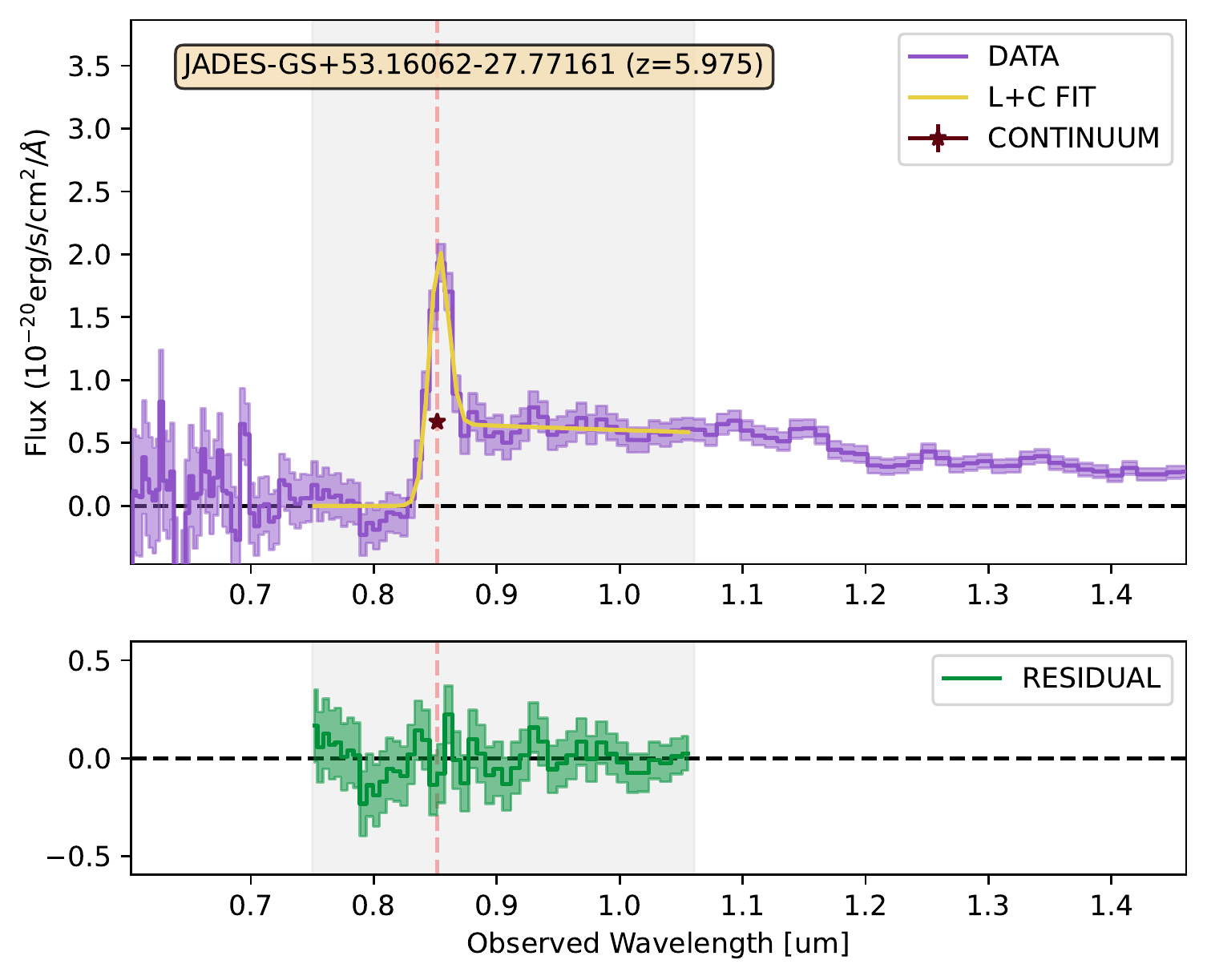}
\includegraphics[width=0.49\textwidth]{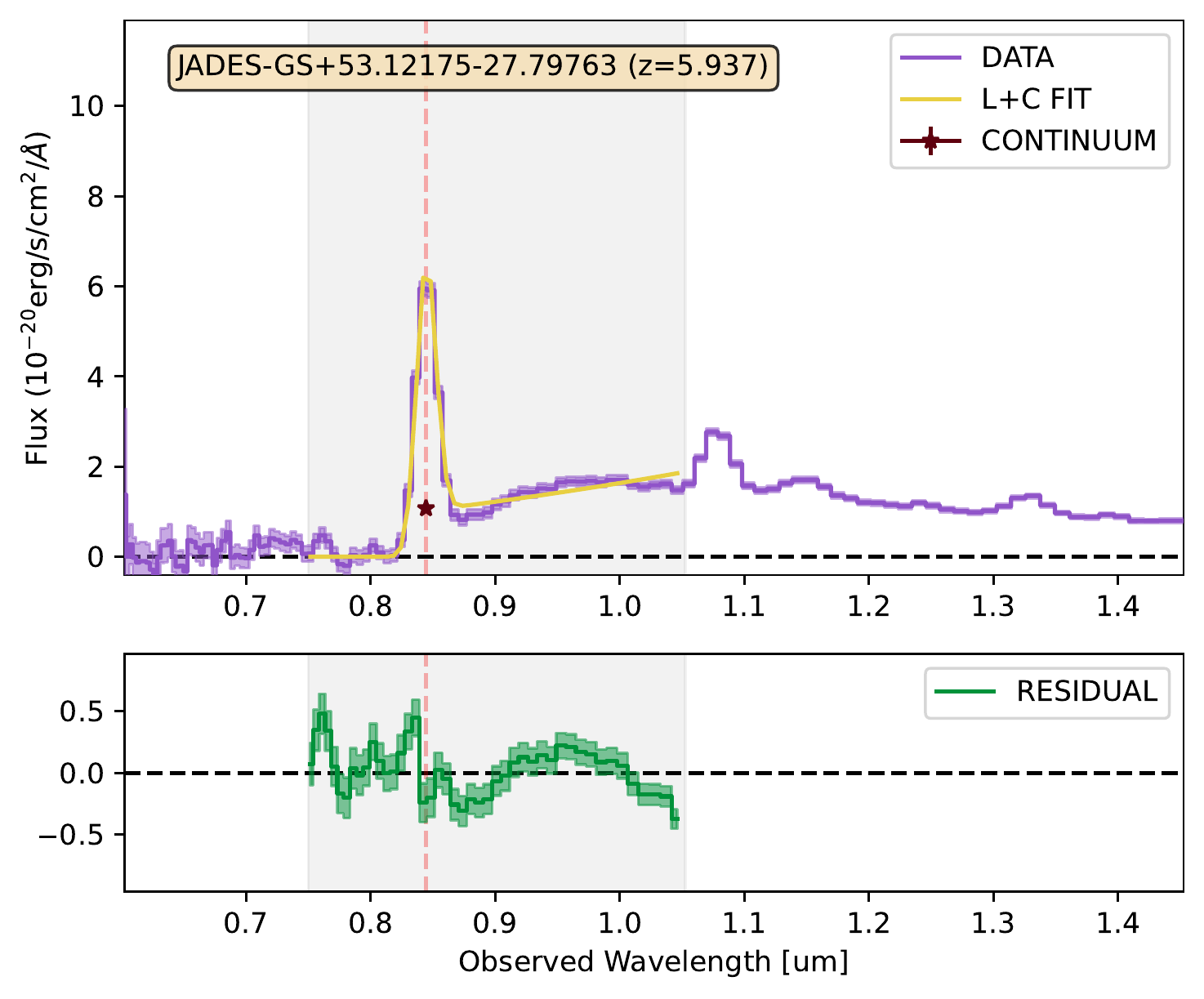}
\includegraphics[width=0.49\textwidth]{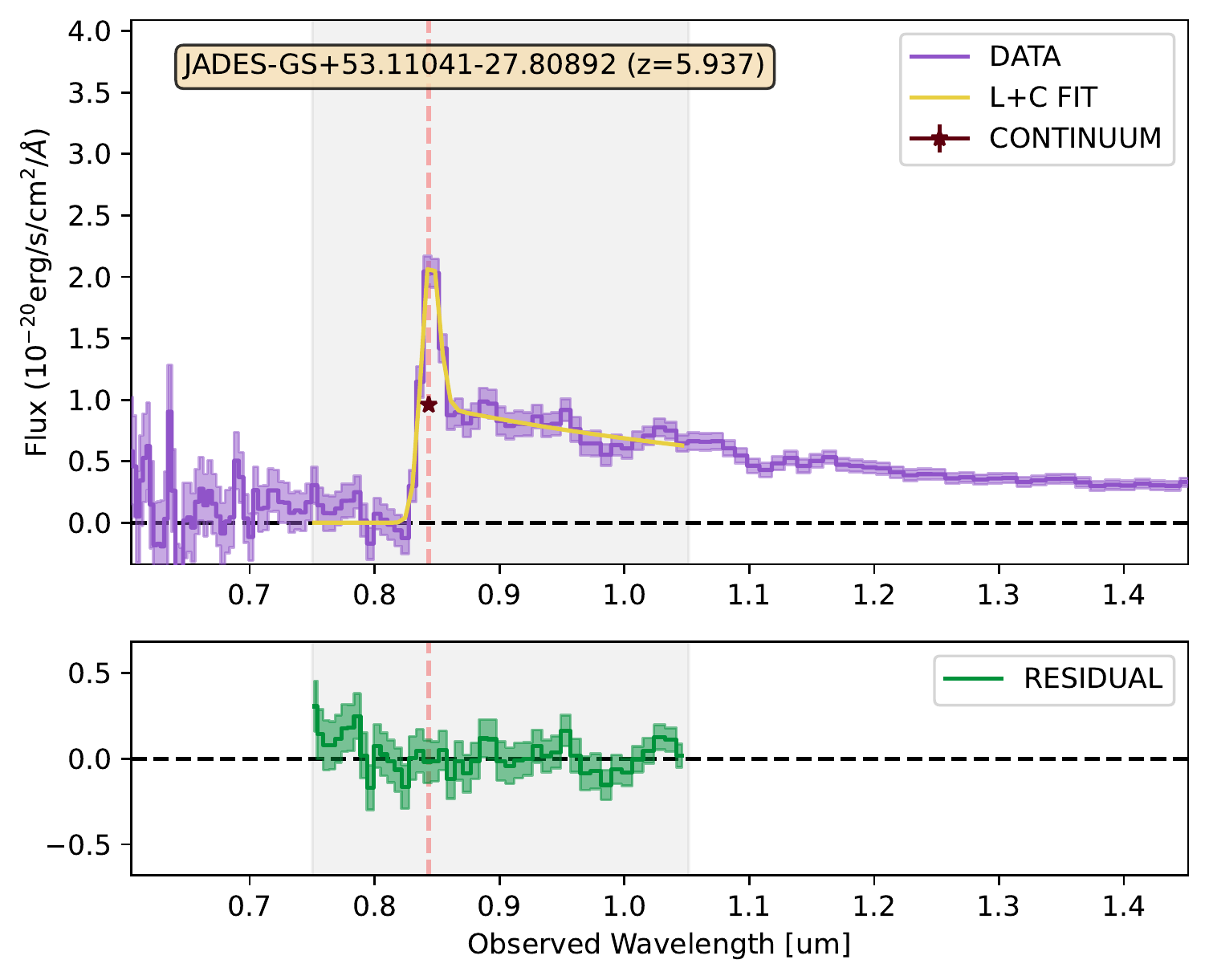}

\caption{See caption of Figure \ref{lyares1}.}
\label{lyares2}
\end{figure*}

\begin{figure*}
\includegraphics[width=0.49\textwidth]{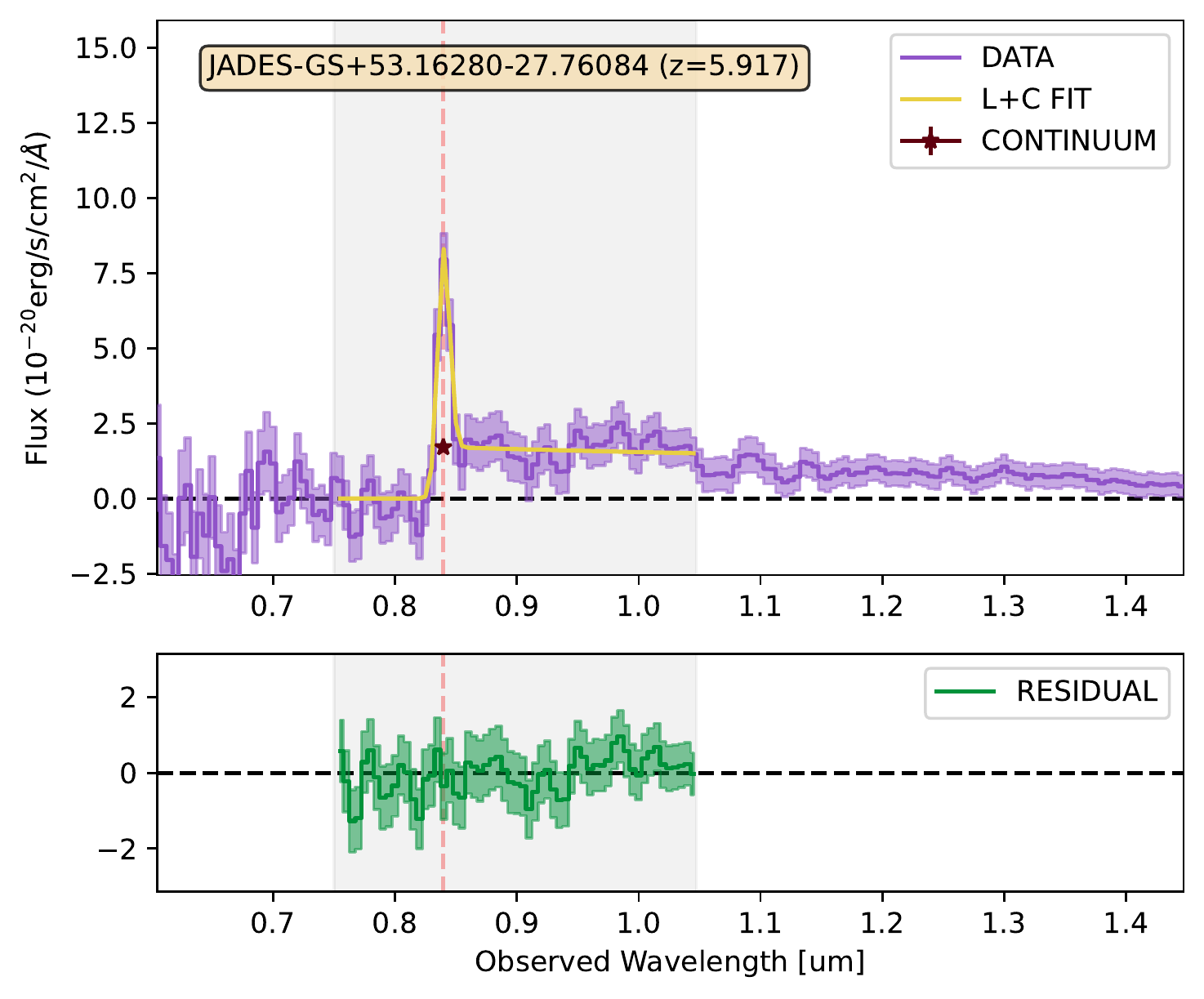}
\includegraphics[width=0.49\textwidth]{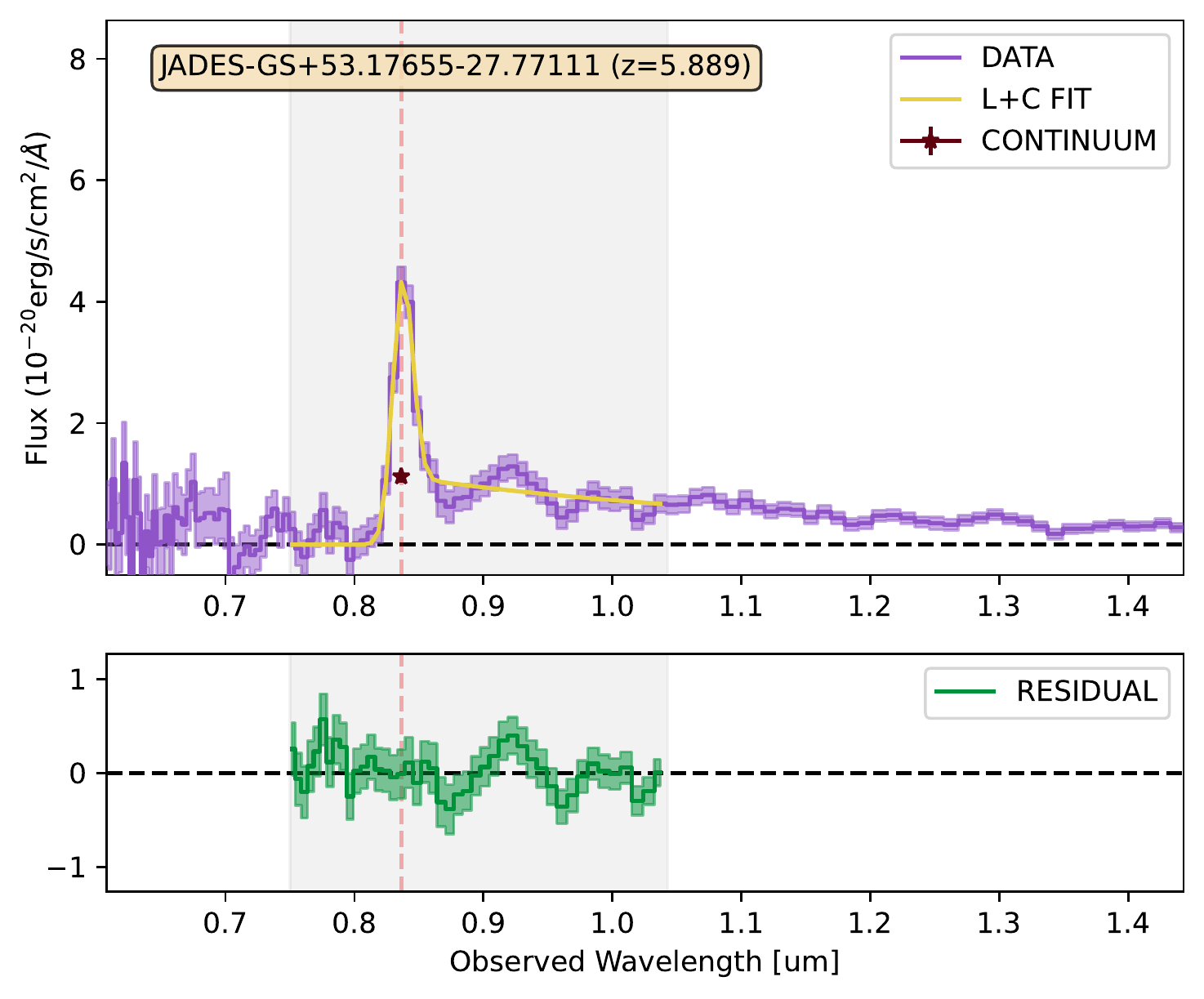}
\includegraphics[width=0.49\textwidth]{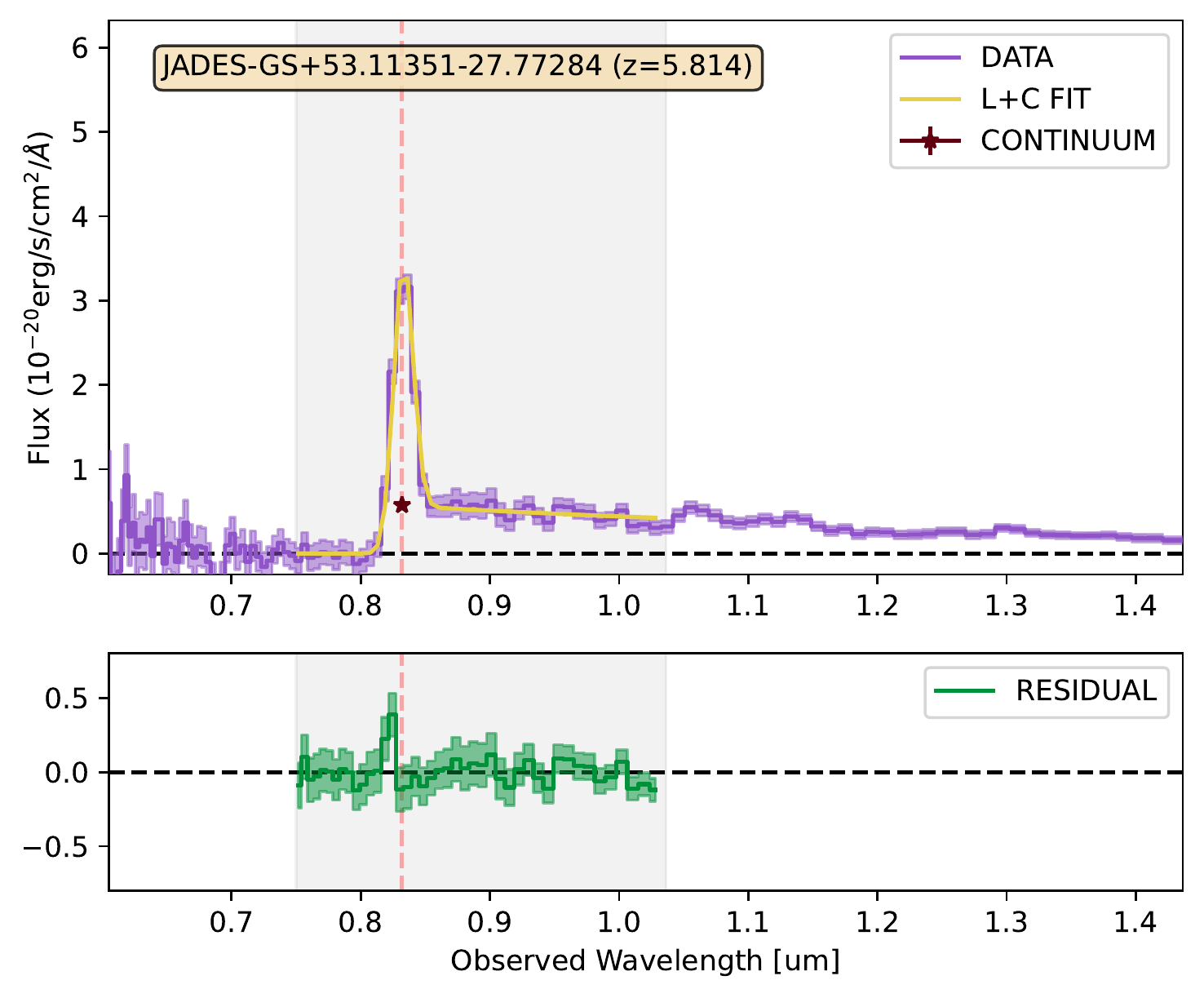}
\includegraphics[width=0.49\textwidth]{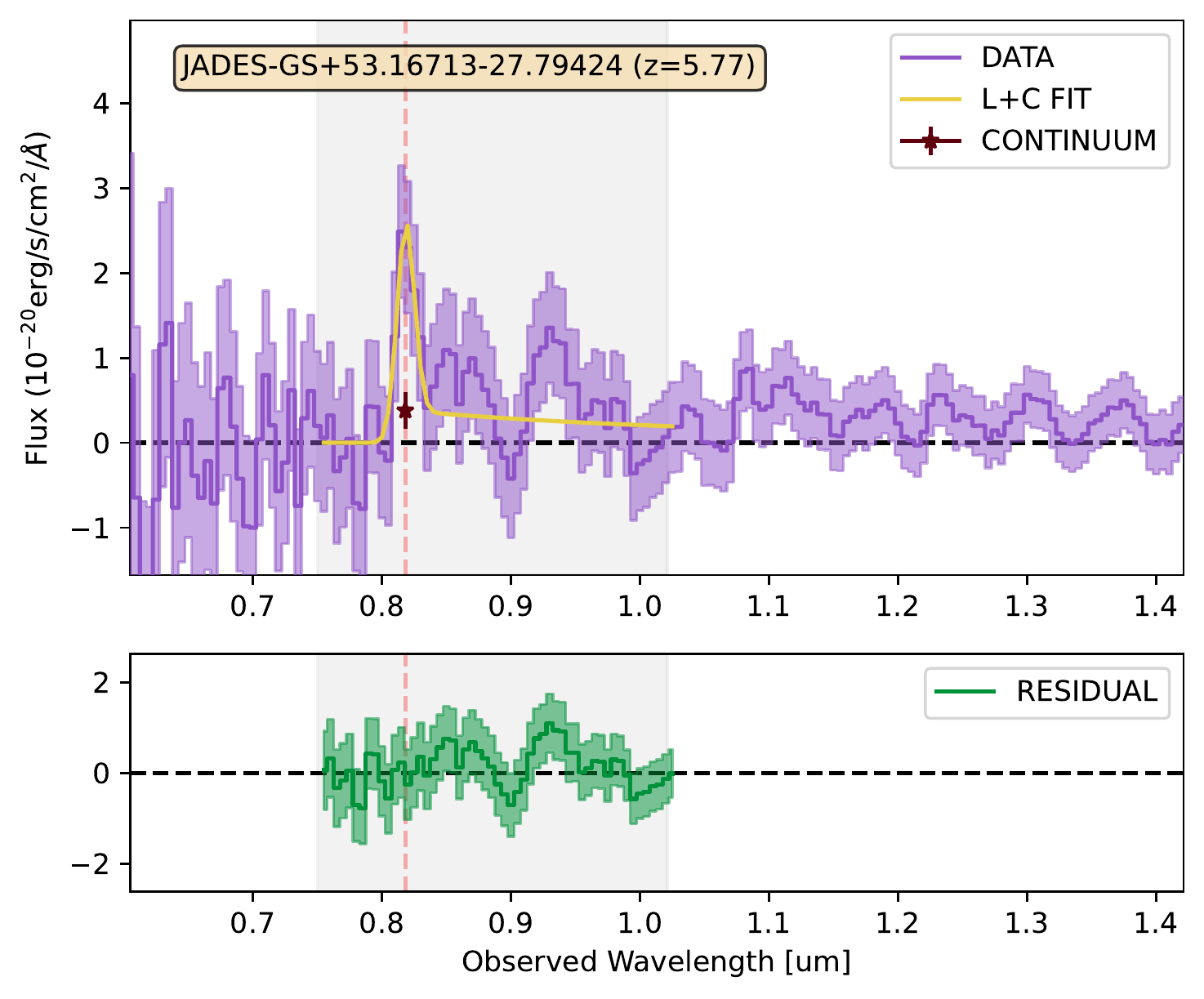}
\includegraphics[width=0.49\textwidth]{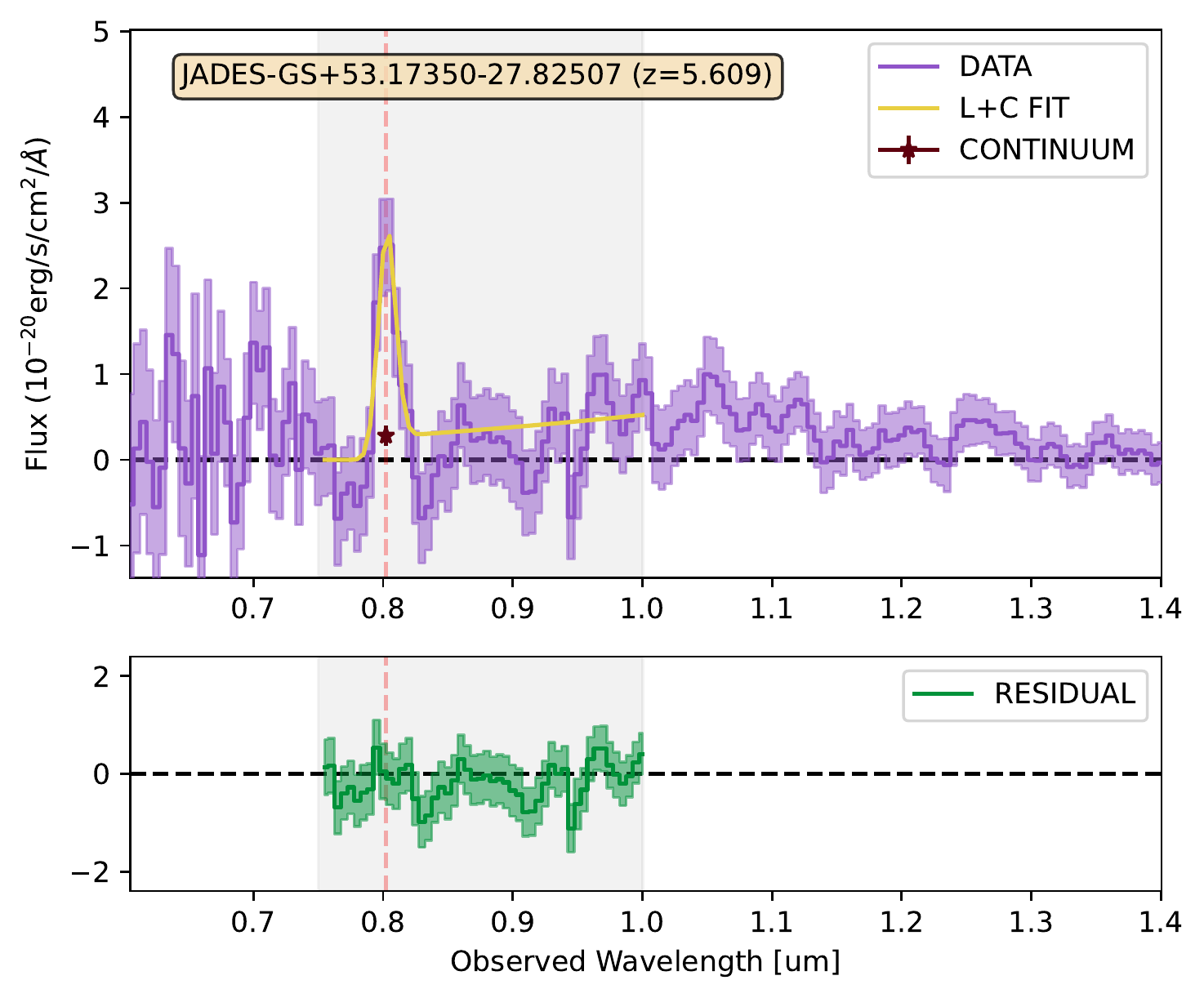}
\caption{See caption of Figure \ref{lyares1}.}
\label{lyares3}
\end{figure*}

With this definition, we find that 17 galaxies in our sample feature significant \lya emission from the R100 spectra alone. The best-fit models of these detections are shown in Figures \ref{lyares1}, \ref{lyares2}, and \ref{lyares3}, while the best-fit parameters of all galaxies in our sample are presented in Table \ref{lyares_table}. 

For comparison, we present \rew values derived by taking the continuum values from this work and the \lya fluxes measured from the R1000 spectra by \citet{saxe23}. These line fluxes were only measured for galaxies at $z>5.8$ in the Deep/HST and Medium/HST tiers in GOODS-S (excluding Medium/JWST, where there are no $z>5.8$ {\refdel galaxies detected in Ly$\alpha$ emission}). For some of these sources, \lya fell into the unobservable chip gap, so no \lya is given. 

We note that the use of the low spectral resolution PRISM/CLEAR grating/filter combination ($R\sim100$) results in detections or limits that are in agreement (i.e., within $3\sigma$) with the higher-resolution R1000 data for every source. This is encouraging, as the latter is more sensitive to low equivalent width lines. For example, \citet{bunk23a} find that \lya is not observable in the R100 spectrum of GNz-11, but is clearly detected in the R1000 spectrum. This can also be seen in our sources that are undetected in the prism, but have a $3\sigma$ \rew$_{,R100}$ upper limit that agrees with a smaller \rew$_{,R1000}$ value (i.e. our \rew upper limit from the prism is consistent with the value of the \rew inferred from the grating detection). 

From our R100 fitting analysis, we find that 17 of the 84 galaxies in our sample are detected in \lya. In the following subsections, we analyze the properties of these detections. 


\clearpage
\onecolumn
\begin{landscape}
\include{LyaOutputTable}
\end{landscape}
\clearpage
\twocolumn

\subsection{Completeness analysis}\label{completeness_sec}
As seen in equation \ref{drew}, our REW sensitivity is dependent on observational parameters (error spectrum and LSF) as well as source properties (redshift and continuum flux). To further complicate matters, the error spectrum features higher values at small wavelengths, resulting in larger uncertainties in $E(\lambda_{Ly\alpha,obs})$ for lower redshift sources. Our sample is quite diverse in redshift, continuum strength (i.e., M$_{\rm UV}$), and sensitivity (i.e., Deep and Medium tiers). So while equation \ref{drew} may be used as a limit on REW, it does not capture the breadth of galaxy properties in our sample, and an estimation of the completeness of our sample is required (e.g., \citealt{thai23}).

To begin, we assume a similar model to the previous subsections: a power-law continuum, a \lya break given by a Heaviside step function, and \lya emission quantified as an REW.  The mean uncertainty spectra for each tier are calculated by averaging the corresponding error spectra. For each galaxy, we take the best-fit continuum strength (M$_{\rm UV}$) and redshift ($z$), and create 50 mock spectra by sampling from uniform distributions of $\beta=[-2.5,2.5]$, and $F_R=[0.1,0.8]$. This process is repeated for three REW values ($25\angstrom$, $50\angstrom$, $75\angstrom$). Gaussian noise is added based on the error spectrum. Each of these 12900 model spectra is fit with the procedure outlined in Section \ref{fitsec}, and the completeness for each galaxy and REW value is then estimated as the fraction of models that are well fit (i.e., that return a REW value within $3\sigma$ of the input value). 

{\refdel This analysis yields an average completeness for our sample of $C_{25\angstrom}=0.33$, $C_{50\angstrom}=0.60$, and $C_{75\angstrom}=0.74$.} As expected, the completeness of each tier increases with REW. Our completeness at REW$=25\angstrom$ is low, which results in poor constraints on $X_{Ly\alpha}$ and X$_{HI}$ (see Section \ref{lyafracsec} and \ref{neufracsec}). We will use these completeness values to derive corrected \lya fractions in the next Section.

\section{Discussion}\label{sec_disc}

\subsection{Source properties}

\begin{figure}
    \centering
    \includegraphics[width=0.49\textwidth]{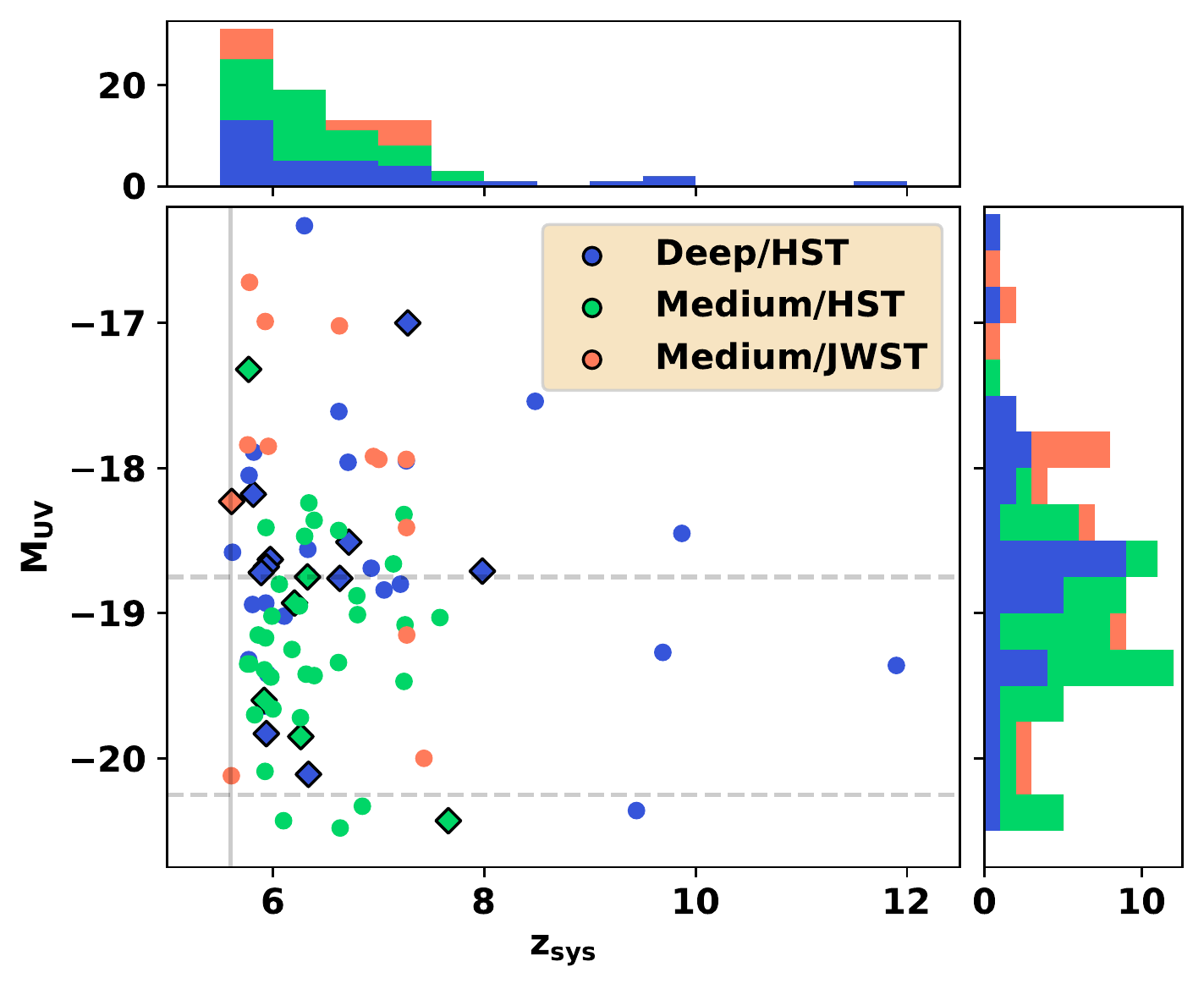}
    \caption{M$_{\rm UV}$ (from NIRSpec spectra, see Appendix \ref{m1500app}) versus systemic redshift (based on optical lines) for our sample. Galaxies observed in different tiers are coloured differently. Sources detected in \lya emission are shown as diamonds with black outlines. Vertical dashed lines show M$_{\rm UV}$ values of -18.75 and -20.25, while the vertical grey line shows our lower redshift cutoff ($z_{sys}>5.6$)}
    \label{fig_zmuv}
\end{figure}

To demonstrate the multi-tier complexity of our sample, we show the systemic redshift (based on the identification of rest-frame optical lines; \citealt{bunk23b}) and the $1500\angstrom$ continuum magnitude (hereafter M$_{\rm UV}$), separated by survey tier (Figure \ref{fig_zmuv}). Previous studies defined UV-faint galaxies as those with $M_{UV}>-20.25$ (e.g., \citealt{curt12}), which shows that most of the JADES sample contains faint sources. The high-redshift bins ($z>8$) are dominated by the Deep/HST sources, but the $5.6<z<7.5$ regime is well explored by all three subsamples. The overall sample is well sampled in the $M_{UV}\sim-19.5$ to $-17.5$ regime, but extends down to $M_{UV}\sim-16.5$.

The sources with \lya detections in R100 data are contained within $z\sim5.6-8.0$, and span a wide range of $M_{UV}\sim-20.5$ to $-17$ . Of the 51\,sources in the Medium tiers, 7 are detected in \lya ($\sim$ 14$\%$). On the other hand, 10 of the 33 Deep galaxies are \lya-detected ($\sim 30\%$). In the next subsection, we will explore the limits that the non-detections imply.

\subsection{Equivalent width - UV magnitude relation}\label{rewmuvsec}
\begin{figure*}
    \centering
    \includegraphics[width=\textwidth]{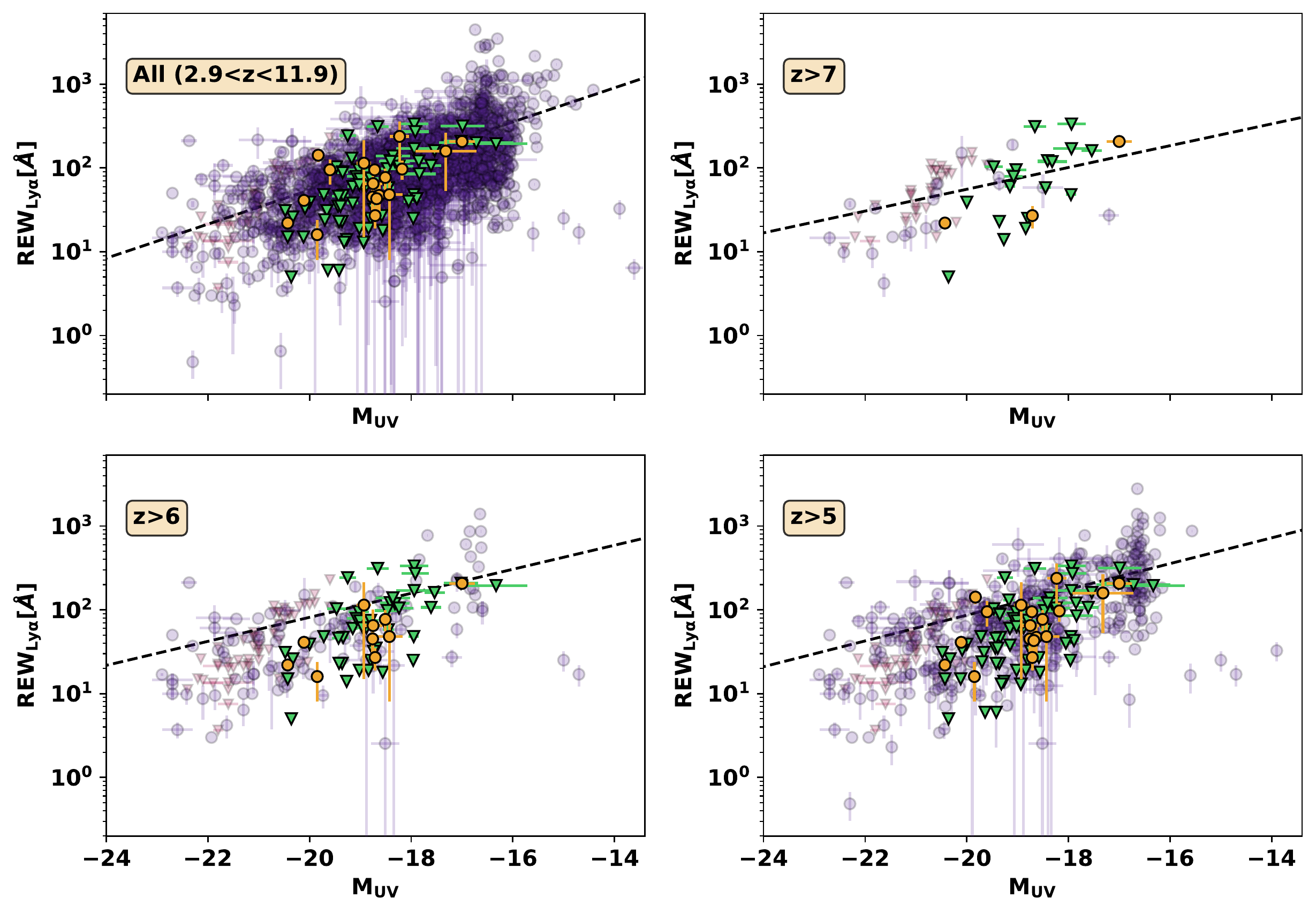}
    \caption{Distribution of rest-frame \lya equivalent widths as a function of M$_{UV}$ for our sample (orange circles for detections and green triangles for $3\sigma$ upper limits) and from literature (purple circles for detections and red triangles for $3\sigma$ upper limits). An illustrative fit to the detections is shown by the black dashed line. See Appendix \ref{compdata} for details of literature sample.}
    \label{fig_evwmuvz}
\end{figure*}

A recent analysis of JWST/NIRSpec MSA data (CEERS; \citealt{tang23}) and data at lower redshift showed a positive correlation between \rew and $M_{UV}$ for a sample of galaxies with high O32 values (i.e., a high level of {\refdel ionisation}). To investigate this relation further, we first collect a literature sample of galaxies with reported spatial positions, spectroscopic redshifts, \rew from spectral observations, and M$_{\rm UV}$ values (including that of \citealt{tang23}; see Appendix \ref{compdata}) and split this sample into different redshift bins (Figure \ref{fig_evwmuvz}). 

In each redshift bin, there appears to be a positive correlation between \rew and M$_{UV}$, such that UV-fainter (higher M$_{UV}$) objects feature higher \lya equivalent widths. To illustrate this, we fit a simple model to the data, resulting in positive slopes (see black dashed line). We do not present the fit values or uncertainty, as the literature sample is not constructed with a single set of criteria. 

While this trend may be physical, it may also be influenced by the sensitivity limits of observations. As a test of this, we create a set of simulated R100 spectra that do not feature any relation between REW and $\rm M_{UV}$, fit them with our method, and plot the resulting best-fit values and upper limits (see Appendix \ref{goodsim}). This test shows that we are not able to recover UV-faint, low-REW galaxies, resulting in an apparent positive correlation. While this does not affect the conclusion of other works, we may not claim a correlation based on our data.

\subsection{Ly$\alpha$ fraction}\label{lyafracsec}

Next, we derive the fraction of galaxies in our sample that are detected in \lya emission ($X_{Ly\alpha}$). This has been a focus of multiple studies over the past decade (e.g., \citealt{star11,curt12,ono12,caru12,caru14,sche14,star17,pent18,yosh22}), where subsamples are usually created according to cuts on M$_{\rm UV}$ and the value of \rew.

The most well-studied sample for $X_{Ly\alpha}$ evolution is that of galaxies with $-21.75<M_{UV}<-20.25$ and \rew$>25\angstrom$ (e.g., \citealt{font10,star11,ono12,curt12,sche12,sche14,pent14,cass15,star17,pent18,yosh22}). For these galaxies, studies have hinted at a steep increase in $X_{Ly\alpha}$ between $z=7-6$, with a more shallow drop off between $z=6-4$. Since the JADES sample contains fainter galaxies (see Figure \ref{fig_zmuv}), we are instead able to focus on fainter galaxies ($\rm M_{UV}>-20.25$).

\begin{figure*}
    \centering
    \includegraphics[width=\textwidth]{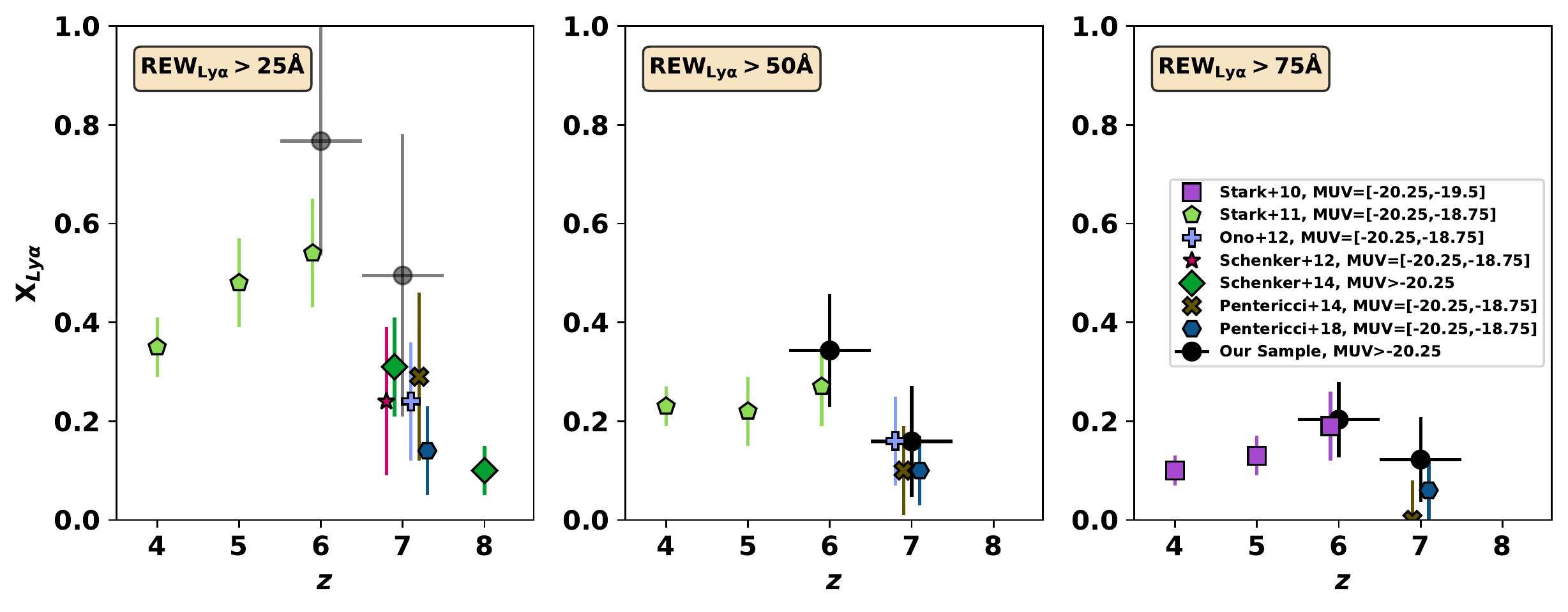}
    \caption{Fraction of observed galaxies detected in \lya emission with REW$_{Ly\alpha}>25\angstrom$ (left), REW$_{Ly\alpha}>50\angstrom$ ({\refdel centre}), and REW$_{Ly\alpha}>75\angstrom$ ({\refdel right}). Derived fractions from literature are shown by coloured markers. For the central panel, note that \citet{star11} and \citet{ono12} used a REW limit of $>55\angstrom$. The fractions derived using only the observed JADES galaxies are shown by black points. Our REW>25$\angstrom$ values (grey) are likely affected by low completeness.
    }
    \label{ewevol}
\end{figure*}

We first divide our sample of galaxies into discrete redshift bins and calculate effective sample sizes by summing their completeness values (Section \ref{completeness_sec}). The \lya fraction is then found by dividing the number of galaxies in the redshift bin who meet the REW limit by the effective sample size (e.g., \citealt{caru12}). This is repeated for three cuts on \rew ($>25\angstrom$, $>50\angstrom$, or $>75\angstrom$), and we compare them to observed fractions from literature for galaxies fainter than $M_{UV}>-20.25$ (\citealt{star10,star11,ono12,sche12,sche14,pent14,pent18}; see Figure \ref{ewevol}). 

For galaxies with $-20.25<M_{UV}<-18.75$, previous studies have shown that the fraction for $EW>25\angstrom$ increases from $z=7$ to $z=6$ with evidence for a decrease to $z=4$, presumably due to the onset of IGM neutrality. Due to the low completeness of our sample at $25\angstrom$ (see {Section \ref{completeness_sec}}), we are not able to place tight constraints on $X_{Ly\alpha}$ in this REW regime. Despite this, our estimates are in agreement with previous findings. The higher completeness at $REW>50\angstrom$ and $REW>75\angstrom$ results in agreement with previous results.

Overall, our sample supports a rise in the \lya fraction between $z=7$ and $z=6$, as seen in previous studies. Since JADES is ongoing, future investigations will include more data and yield tighter constraints on the evolution of this quantity. Even with our current data, we are able to constrain the IGM neutral fraction, as seen in the next subsection.

\subsection{Constraints on neutral fraction}\label{neufracsec}

The observed \lya fraction of galaxies for a given redshift, M$_{\rm UV}$ bin, and \rew limit provides valuable information on the neutral fraction of the IGM (X$_{HI}$). This is due to the fact that the evolution of X$_{Ly\alpha}(z\lesssim6)$ is dependent only on galaxy properties, while at $z\gtrsim6$ it is also dependent on the properties of IGM transmission. Some studies compared observed X$_{Ly\alpha}(z\sim7)$ with the fraction expected from simulations, resulting in a range of estimates for $X_{HI}(z=7)$: $\lesssim0.3$ (\citealt{star10}), $\sim0.5$ (\citealt{caru14}), $\gtrsim0.51$ (\citealt{pent14}), $\sim0.6-0.9$ (\citealt{ono12}), $\lesssim0.7$ (\citealt{furu16}). This discrepancy may be partially explained by sample properties (e.g., difference in $M_{UV}$ ranges and small sample sizes).

The conversion from X$_{Ly\alpha}$ to $X_{HI}$ is nontrivial, and is dependent on the simulation used for comparison to observations. For example, the semi-numerical code DexM (\citealt{mesi07,mesi11,zahn11}) has been used by some works (e.g., \citealt{dijk11,pent14}) to create three-dimensional models of galaxy halos, determine how they {\refdel ionise} their surroundings, and {\refdel characterise} their redshift evolution. The outputs of this process (e.g., $X_{Ly\alpha}$) may then be compared to observations. 

\begin{figure}
\centering
\includegraphics[width=0.5\textwidth]{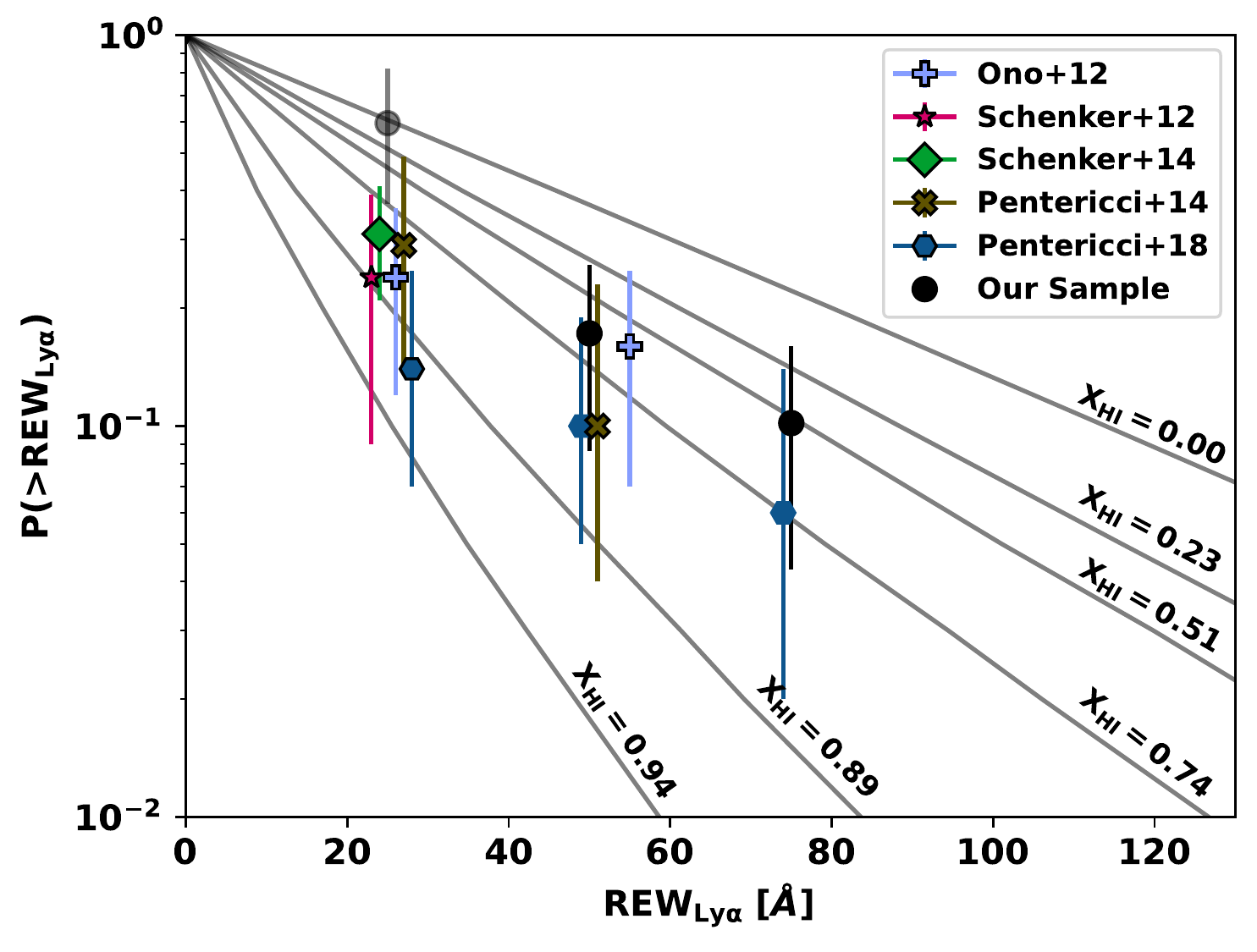}
\caption{Cumulative distribution for \lya rest {\refdel R}EW for UV-faint ($-20.25<M_{UV}<-18.75$) galaxies at $z\sim7$. Each solid line shows the expected distribution for a model with $N_{HI}=10^{20}$\,cm$^{-2}$ and a wind speed of 200\,km\,s$^{-1}$, but with a different neutral fraction \citep{pent14}. Estimates from literature (\citealt{ono12,sche12,sche14,pent14,pent18}) are shifted by $1\angstrom$ for visibility. Our P(REW>25$\angstrom$) point (grey) is likely affected by low completeness.
}
\label{xhiplot}
\end{figure}

While an updated simulation is beyond the scope of this work, we may use our $X_{Ly\alpha}$(REW$_{Ly\alpha}$) values at $z\sim7$ to generate a cumulative distribution {\refdel Function} (CDF) of REW$_{Ly\alpha}$, and compare this to model outputs of \citet{pent14}. This model is appropriate for galaxies with $-20.25<M_{UV}<-18.75$ and assumes $N_{HI}=10^{20}$\,cm$^{-2}$ and a wind speed of 200\,km\,s$^{-1}$, but with a variable neutral fraction. It is based on the assumption that the \lya CDF at $z=6$ and $z=7$  are intrinsically the same, but are observed to differ because of neutral IGM attenuation at $z=7$. As seen in Figure \ref{xhiplot}, higher values of $X_{HI}$ result in less \lya transmission, and thus a steeper CDF. Literature values (\citealt{ono12,sche12,pent14,pent18}) appear to argue for a value of $X_{HI}\sim0.5-0.9$.

Our results (using a wide redshift bin of $6<z<8$) are in agreement with those of the previous studies (e.g., \citealt{maso18}), suggesting an approximate $X_{HI}$ of $\sim0.2-0.7$. Note that our $P(REW>25\angstrom)$ point is influenced by poor completeness, and thus does not provide a constraint. We note that this M$_{UV}$ range is not {\refdel optimised} for our sample, and a future work will investigate how the redshift evolution ($z=8-6$) of $P(>REW)$ for $-20.4<M_{UV}<-16$ galaxies may be used to constrain $X_{HI}(z)$. 

The $X_{HI}$ range of our analysis agrees with previous analyses that use the \rew cumulative distribution (e.g., \citealt{sche14,pent14,pent18}) as well as simulations (e.g., \citealt{maso18}). This suggests that our estimates of $P(>REW)$ have not been underestimated due to the NIRSpec MSA shutters ($0.2''\sim1$\,kpc at $z\sim7$) missing \lya flux from extended {\refdel halos}, as noted by \citet{jung23}. In this previous study, $F_{Ly\alpha,MSA}$ for one observed source was only $20\%$ of the value as derived by MOSFIRE observations. On the other hand, \citet{tang23} found agreement between MSA- and ground-based estimates of $F_{Ly\alpha}$ for four $z\sim7-9$ galaxies (but with large uncertainties for two sources). In addition, large \lya halos are commonly seen at low-$z$, but will not have time to evolve for high-redshift sources. So while slit losses are unlikely to affect our results, this effect may be further investigated by performing NIRSpec/IFU (field of view $3''\times3''$) observations of representative sources, or forward modelling the slit losses in simulations.

\section{Conclusions}\label{sec_conc}
In this work, we present the first constraints on \lya emission using JWST/NIRSpec MSA R100 spectra from the JADES survey. The increased sensitivity of this instrument enables deeper investigations of faint galaxies in the early Universe. Our sample consists of 84 galaxies at $z>5.6$, each with secure spectroscopic redshifts (\citealt{bunk23b}). While the sample is concentrated at $5.6\lesssim z\lesssim7.5$, we include sources up to $z\sim12$. In addition, the M$_{\rm UV}\sim-19.5$ to $-17.5$ range is well probed, but we include sources that are fainter (M$_{\rm UV}\sim -16$) and brighter (M$_{\rm UV}\sim -20.4$)

By fitting each spectrum with a line and/or continuum model, accounting for the spectral dispersion, and comparing the relative goodness of fit values, we find that 17 sources at $z\sim5.6-8.0$ show evidence for \lya emission in R100. The strong continuum of each source enables us to estimate the continuum flux (and M$_{\rm UV}$) at the \lya wavelength directly from the spectra. We derive \lya rest-frame equivalent widths for each source.

We build a large comparison sample from literature of galaxies with estimates of spectroscopic redshifts, M$_{\rm UV}$, and \rew. By combining the JADES and literature samples, we find that the reported positive correlation between M$_{\rm UV}$ and \rew is supported for galaxies in multiple redshift bins, but the observed correlation in our data may be caused by sensitivity effects. 

Next, we calculate the redshift evolution of the \lya fraction ($X_{Ly\alpha}$) in bins of \rew. Due to the faintness of the JADES sample (M$_{\rm UV}\gtrsim-20.4$), we are able to place constraints on the poorly studied faint, high-redshift ($z\sim6-7$) evolution of this fraction: a shallow increase from $z=7$ to $z=6$ for REW$>50\angstrom$ and REW$>75\angstrom$.

The distribution of \rew values was then used to place a constraint on the neutral fraction ($X_{HI}$) at $z\sim7$ using the model of \citet{pent14}. Our results indicate $X_{HI}\sim$ 0.2-0.7, which is in agreement with previous studies. 

The JADES survey is still ongoing, so this dataset will expand with time. In addition, many sources feature higher resolution R1000 spectra, which enable further science cases such as \lya velocity offset and line asymmetry analysis, damping wing modelling, and environments of \lya emitters. Combined, these analyses will reveal the details of reionization in unprecented detail.

\section*{Acknowledgements}
GCJ, AJB, AS, AJC, and JC acknowledge funding from the ``FirstGalaxies'' Advanced Grant from the European Research Council (ERC) under the European Union’s Horizon 2020 research and innovation programme (Grant agreement No. 789056).
JW, RM, WMB, TJL, LS, and JS acknowledge support by the Science and Technology Facilities Council (STFC) and by the ERC through Advanced Grant 695671 ``QUENCH''.
JW also acknowledges funding from the Fondation MERAC.
RM also acknowledges funding from the UKRI Frontier Research grant RISEandFALL and from a research professorship from the Royal Society.
SA acknowledges support from Grant PID2021-127718NB-I00 funded by the Spanish Ministry of Science and Innovation/State Agency of Research (MICIN/AEI/ 10.13039/501100011033).
RB acknowledges support from an STFC Ernest Rutherford Fellowship (ST/T003596/1).
This research is supported in part by the Australian Research Council Centre of Excellence for All Sky Astrophysics in 3 Dimensions (ASTRO 3D), through project number CE170100013.
SC acknowledges support by European Union’s HE ERC Starting Grant No. 101040227 - WINGS.
ECL acknowledges support of an STFC Webb Fellowship (ST/W001438/1).
DJE, BDJ, and BER acknowledge a JWST/NIRCam contract to the University of Arizona NAS5-02015.
DJE is supported as a Simons Investigator.
Funding for this research was provided by the Johns Hopkins University, Institute for Data Intensive Engineering and Science (IDIES).
RS acknowledges support from a STFC Ernest Rutherford Fellowship (ST/S004831/1).
H\"{U} gratefully acknowledges support by the Isaac Newton Trust and by the Kavli Foundation through a Newton-Kavli Junior Fellowship.
The research of CCW is supported by NOIRLab, which is managed by the Association of Universities for Research in Astronomy (AURA) under a cooperative agreement with the National Science Foundation.
GCJ would like to thank C. Witten and N. Laporte for valuable insight into the sample.
We thank the anonymous referee for constructive feedback that has enhanced this work.

\bibliographystyle{aa}
\bibliography{references}

\begin{appendix}

\section{Damping wing in R100}\label{dampapp}
In this work, we assume that the Lyman break may be approximated by a step function. To test whether this is appropriate for our R100 spectra, we consider the wavelength-dependent damping wing optical depth formalism of \cite{mira98} for a source at redshift $z_{s}$:
\begin{equation}
\tau(\Delta\lambda)=\frac{\tau_oR_{\alpha}}{\pi}(1+\delta)^{3/2}\left[F(x_2)-F(x_1)\right]    
\end{equation}
where $\Delta\lambda$ is the wavelength offset from the redshifted centroid of Ly$\alpha$ ($\lambda_{\alpha}$), $\delta\equiv\Delta\lambda/[\lambda_{\alpha}(1+z_s)]$, and we assume $\tau_o=\tau_{GP}=(7.16\times10^5)\times\left(\frac{1+z_{s}}{10}\right)^{3/2}$ and $R_{\alpha}=2.0136\times10^{-8}$ from \citet{mesi08}. F(x) is given as:
\begin{equation}
F(x)=\frac{x^{9/2}}{1-x}+\frac{9x^{7/2}}{7}+\frac{9x^{5/2}}{5}+3x^{3/2}+9x^{1/2}-\frac{9}{2}\mathrm{log}\frac{1+x^{1/2}}{1-x^{1/2}}
\end{equation}
with $x_1=(1+z_n)/[(1+z_s)(1+\delta)]$ and $x_2=1/(1+\delta)$ where $z_n$ is the redshift where absorption by the IGM is assumed to be negligible (we use the standard assumption of $z_n=6${\refdel; e.g., }\citealt{mort16,fan22}). We note that in this form, the model assumes a uniform $X_{HI}$ between $z_n<z<z_s$ (here assumed to be unity) and a no IGM absorption below $z<z_n$.

As $z_s$ approaches $z_n$ from high values, the damping wing begins to approximate a step function. For sources below $z_n$ (which includes most sources in our sample), the IGM is expected to have little effect, and a step function is thus appropriate. But for sources above $z_n$, it is possible that the damping wing will have an effect.

\begin{figure}
\centering
\includegraphics[width=0.5\textwidth]{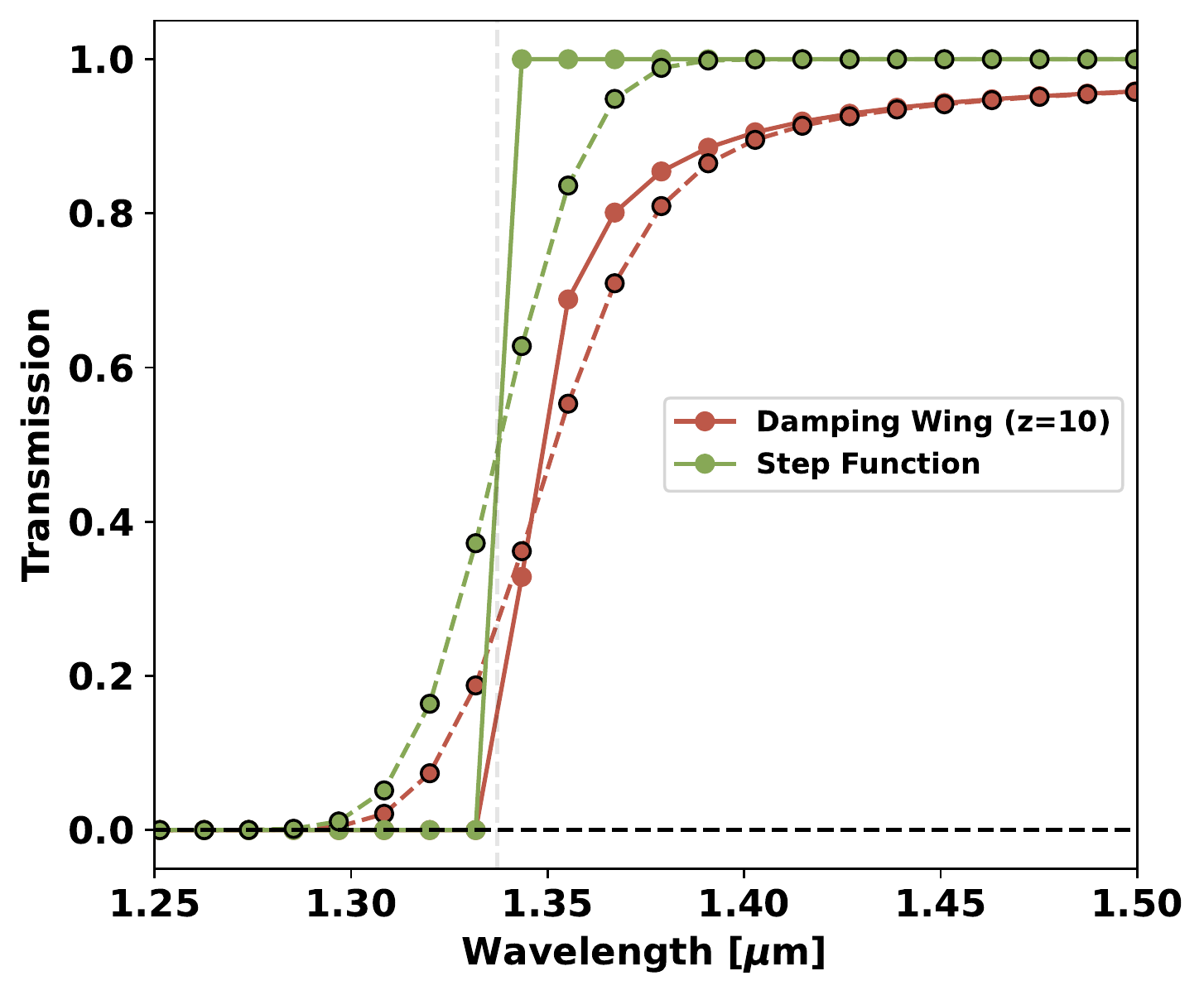}
\caption{Transmission models of a damping wing (brown lines) and a step function (green lines) for a $z=10$ source. We include the intrinsic model, regridded to match our R100 observations (solid lines), and the dispersed version of this model (dashed lines). The \lya wavelength is shown by a faint vertical line.}
\label{damp}
\end{figure}

To investigate this, the optical depth model is used to create a transmission spectrum for a source at $z_s=10$, using the same wavelength grid as our R100 spectra (see solid lines of Figure \ref{damp}). To account for the instrumental dispersion, this model is convolved with a Gaussian with $\sigma=\sigma_R$ (see Section \ref{reseff}; dashed lines of Figure \ref{damp}). 

For $\lambda>>\lambda_{Ly\alpha}$ or $\lambda<<\lambda_{Ly\alpha}$, the two convolved curves are similar (i.e., either 0 or unity). But in this case of a $z=10$ source, the curves differ at  $\lambda\sim\lambda_{Ly\alpha}$, with a discrepancy of up to $\sim20\%$. This strong difference may be detected for some sources, and will be used in future works to place constraints on the proximity zones and IGM neutral fraction of $z>8$ sources in JADES (e.g., Jakobsen et al. in prep). Indeed, multiple studies have now used JWST/NIRSpec observations to constrain the \lya damping wing (e.g., \citealt{fuji23,hein23,umed23}).

Throughout this work, we assume that the transmission function for galaxies in our sample is a simple step function (i.e., $100\%$ transmission redwards of $\lambda_{Ly\alpha}$). In reality, damping wings will affect all of the $z\gtrsim8$ sources, resulting in low transmission at $\lambda_{Ly\alpha}$ and thus making it more difficult to detect \lya emission. For the six $z>8$ sources in our sample (none of which are detected in \lya emission), we note that this is not accounted for in our \rew upper limits, so these may be slightly underestimated. But a full treatment of the damping wing effect is beyond the scope of this paper and will be fully explored in a future analysis.

\section{M$_{\rm UV}$ Derivation}\label{m1500app}

Due to the high quality of the NIRSpec spectra, we are able to derive M$_{\rm 1500\angstrom}$ absolute magnitudes (M$_{\rm UV}$) directly from the observed data. First, the observed data are shifted from the observed to rest frame by multiplying all flux values (i.e., $f_{\lambda}$) by $1+z$ and dividing all wavelength values by the same factor. The rest-frame $f_{\lambda}$ values are converted to a $f_{\nu}$:
\begin{equation}
f_{\nu}=f_{\lambda}\lambda^2/c
\end{equation}
where $c$ is the speed of light. We collect and average all $f_{\nu}$ values that lie between $1400-$1600\,\angstrom (rest-frame), and use this average value ($\bar{f_{\nu}}$) to derive an apparent AB magnitude (\citealt{oke83}):
\begin{equation}
m_{1500}=-2.5\log_{10}(\bar{f_{\nu}})-48.60
\end{equation}
which is converted to an absolute magnitude:
\begin{equation}
M_{\rm UV}=m_{1500}-5\log_{10}(D_L[Mpc])-25
\end{equation}

An error is estimated by calculating the {\refdel root mean square} noise level of the $f_{\nu}$ values between $1400-$1600\,\angstrom from the error spectrum and perturbing $\bar{f_{\nu}}$ by this value.

\section{Comparison data}\label{compdata}
As one of the brightest emission lines for star-forming galaxies at high-redshift, \lya has been studied in numerous galaxies over the past decades. While the subset of JADES that we {\refdel analyse} in this work offers the opportunity to explore the \lya properties of galaxies between $z\sim5-11$, our conclusions are strengthened by the addition of archival data. Here, we collect a large literature sample of galaxies with reported spatial positions, spectroscopic redshifts, \lya equivalent widths from spectral observations, and M$_{UV}$ values. Unless otherwise stated, $z_{sys}=z_{Ly\alpha}$ and M$_{UV}=M_{1500}$. The details of our comparison sample are given below, and we present a full machine readable table\footnote{(See published version for permanent CDS link) \url{https://drive.google.com/file/d/1idNB3hEUXLCg4JXPQjv4SjS9KtzdI_dS/view?usp=sharing}}. To avoid repeated galaxies, we search for entries within $0.25''$ of each other and exclude the older measurement.

Followup spectroscopy of bright {\refdel Ly-$\alpha$ emitters (}LAEs{\refdel )} discovered in the Systematic Identification of LAEs for Visible Exploration and Reionization Research Using Subaru HSC (SILVERRUSH) with a variety of ground-based telescopes resulted in numerous detections (\citealt{shib18}). The continuum level underlying \lya was found by extrapolating from red filters ($\beta=-2$), while M$_{UV}$ was estimated from the observed spectra.

We also include the large survey CANDELSz7 \citep{pent18}, a large program that observed {\refdel star-forming galaxies} at $z\sim6$ and $z\sim7$ in the GOODS-South, UDS, and COSMOS fields with the VLT/FORS2 spectrograph. We include all galaxies with good quality flags (i.e., A, A/B, and B) and \lya flux estimates. M$_{UV}$ values are estimated using fits to CANDELS photometry, while the REW is derived by fitting the observed spectra. Six sources are also observed with MUSE by \citet{keru22}, so they are not included.

Next, we include the results of studies that used Keck/MOSFIRE. \citet{jung22} observed eight $z\sim7-8$ galaxies in CANDELS EGS and used an asymmetric Gaussian to fit the \lya emission. We exclude one source that lacks a spectroscopic redshift and a number of galaxies with only upper limits on \rew. Four galaxies in this sample likely re-observed in \citet{tang23}, so they are also excluded. While \citet{tilv20} observed three galaxies in group at $z=7.7$, we take the one source with M$_{UV}$ calculated by \citet{tang23}, or $z8\_5$ (EGS-zs8-1 in \citealt{oesc15}). \citet{song16} detected one galaxy in \lya emission ($z7\_GSD\_3811$), whose emission was fit with an asymmetric profile. \citet{hoag19} detected \lya emission from two galaxies that are strongly lensed by galaxy clusters. Values are magnification-corrected, and equivalent width is calculated by dividing the line flux by HST WFC3/F160W continuum flux.

Keck/DEIMOS spectroscopy by \citet{ono12} resulted in line detections for three $z\sim7$ galaxies in the SDF and GOODS-North. The continuum properties were derived from a fit to the photometry. Further Keck/DEIMOS observations revealed 36 LAEs (\citealt{full20}). Note that SDF-63544 is also known as IOK-1 (e.g., \citealt{iye06}). Additional Keck observations resulted in three detections of \lya with the spectrographs LRIS and NIRSPEC (\citealt{sche12}), where the continuum was estimated by extrapolating a power-law model ($\beta=-2$) from a red filter.

Two works used the VLT/FORS2 spectrograph (\citealt{cuby03,vanz11}). For these, we take M$_{UV}$ from the compilation of \citet{matt19}. Additionally, the GMOS spectrographs on the 8.2m Gemini Telescopes were used to observe one target. The resulting spectra were fit with templates, resulting in a \lya REW and M$_{1350}$.

The MMT/Binospec spectrograph was used to observe eight UV-bright (M$_{\rm UV}\sim-22$) galaxies at $z\sim7$ selected from the ALMA REBELS survey (\citealt{ends22}). Red bands were used to estimate the \lya continuum, while M$_{UV}\equiv M_{1600}$.

VLT/MUSE features a bluer spectral range ($0.465-0.930\,\mu$m\footnote{\url{https://www.eso.org/sci/facilities/paranal/instruments/muse/inst.html}}) with respect to the JWST/NIRSpec PRISM/CLEAR filter/disperser combination ($0.6-5.3$\,$\mu$m), allowing it to probe \lya to lower redshifts ($z\sim2.8-6.7$). Through the MUSE-WIDE and MUSE-DEEP surveys, \citet{keru22} present the \lya equivalent widths for 1920 galaxies over the full redshift range accessible to MUSE. The continuum level at $\lambda_{Ly\alpha}$ is estimated from photometry, while \lya is measured from each MUSE data cube.

We also include 10\,galaxies in the Abell 2744 cluster from the recent work of \citet{prie23}, who {\refdel analysed} data from the GLASS-JWST {\refdel Early Release Science} program. M$_{UV}$ values were derived from SED fits to HST and JWST photometry, while REW$_{Ly\alpha}$ is taken from the MUSE observations of \citet{rich21}. The systemic redshift is derived from fits to optical lines as observed by JWST.

Finally, we include results from the recent work of \citet{tang23}, who used the MSA of JWST/NIRSpec (R100 and R1000) to extract spectra and continuum estimates for ten sources. Four {\refdel sources} feature small spatial offsets ($<0.1''$) and redshift differences ($\delta z\sim0.25-0.35$) with sources observed with Keck/MOSFIRE by \citet{tang23}. Since the redshift difference may be explained by a difference in methods (i.e., photometric vs. spectroscopic), we assume that these are the same galaxies. Here, $z_{sys}=z_{[OIII]5007}$. \lya and continuum properties were extracted from the {\refdel CEERS} spectra and from photometry, respectively. The NIRSpec-based \lya properties agree with ground-based REW measurements, suggesting the absence of calibration issues. A few of these sources were known by other names in previous studies: CEERS-1019 is EGSY8p7 (\citealt{zitr15}), CEERS-1029 is EGS\_z910\_44164 (\citealt{lars22}), and CEERS-698 is EGS-zs8-2 (\citealt{robe16}).

\section{REW-M$_{\rm UV}$ simulation}\label{goodsim}
In Section \ref{rewmuvsec}, we found that the \rew and M$_{\rm UV}$ of our sample showed an apparent positive relation, such that more UV-faint galaxies featured larger \lya equivalent widths. But since REW is a ratio of line flux to continuum flux, it is possible that this relation is influenced by the sensitivity limit of our sample and fitting method (i.e., we may miss \lya-faint galaxies). In this Section, we test this possibility using a method similar to our completeness analysis (Section \ref{completeness_sec}).

For each survey tier (Deep/HST, Medium/HST, Medium/JWST), we create 750 model spectra by sampling uniformly for several properties ($5.6<z<12.0$, $-20.4<M_{UV}<-16.4$, spectral slope $-2<n<2$, $0.1<F_R<0.8$) and log-uniformly for \lya equivalent width ($1<$REW$_{Ly\alpha}<750\angstrom$). Gaussian noise is added to each model spectrum based on the corresponding mean error spectrum, and the resulting spectrum is fit with our routine. 

\begin{figure}
    \centering
    \includegraphics[width=0.5\textwidth]{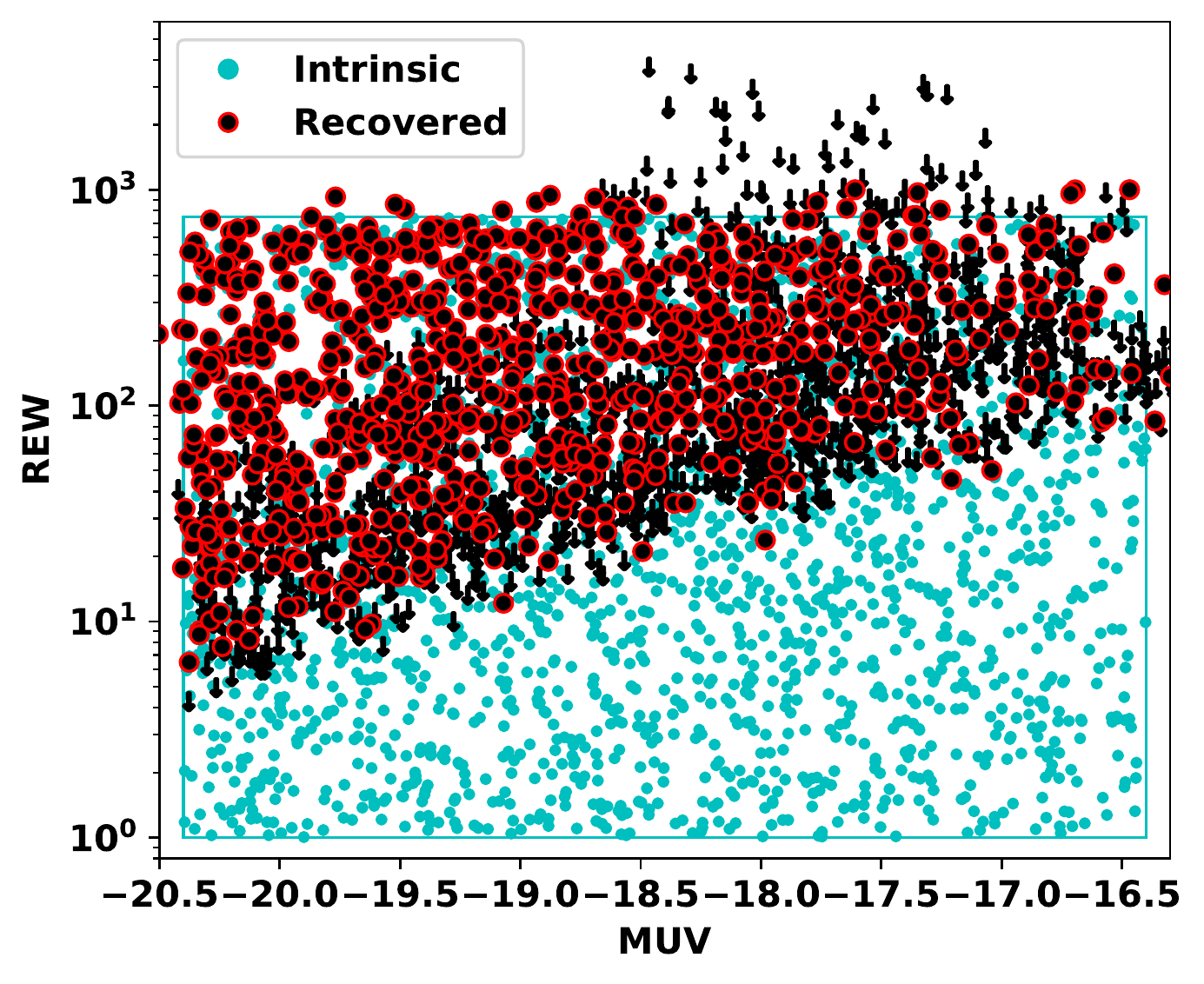}
    \caption{Results of fitting a sample of simulated galaxies. The intrinsic distribution of REW and M$_{\rm UV}$ for the simulated galaxies are shown as cyan circles, with a cyan rectangle outlining the region. The best-fit values and $3\sigma$ upper limits are shown as red-outlined circles and black arrows, respectively.}
    \label{rewmuvsim}
\end{figure}

In Figure \ref{rewmuvsim}, we show the intrinsic distribution of REW vs M$_{UV}$ in blue, and display the resulting best-fit values as red-outlined black circles. Upper limits on REW ($3\sigma$) are shown as downwards arrows. It is clear that we do not capture the full distribution, as galaxies in the lower-right corner (UV-faint and low-REW) are not able to be fit. From this test, it is clear that our sample and fitting approach may result in a false positive trend of REW with M$_{UV}$, due to sensitivity limits.

\end{appendix}

\label{lastpage}
\end{document}

%% file: LyaOutputTable.tex
\begin{longtable}{ccc|cc|cc|cc}
\caption{Results of fitting R100 spectra of each galaxy. We include the best-fit $\rm M_{UV}$ (from the spectrum, see Appendix \ref{goodsim}) and continuum flux at the redshifted \lya wavelength. Line fluxes and rest-frame equivalent widths from the R100 data alone {are shown first. Next, } REW values derived by combining the R1000-based \lya fluxes from \citet{saxe23} with our R100 continuum values are presented for reference. Systemic redshifts (based on fits to strong optical lines) have uncertainty of $0.001$ \citep{bunk23b}. Upper limits are given as $3\sigma$. Galaxies detected in \lya emission in the R100 data are listed first.\label{lyares_table}}
\\ID & $z_{sys}$ & Tier & M$_{\rm UV}$ & $S_{C}(\lambda_{Ly\alpha,obs})$                                               & $F_{Ly\alpha,R100}$                     & $REW_{Ly\alpha,R100}$ & $F_{Ly\alpha,R100}$                     & $REW_{Ly\alpha,R100}$\\ 
\textit{JADES-GS}   &	        &	   & 			   & $10^{-21}$\,erg\,s$^{-1}$\,cm$^{-2}$\,$\angstrom^{-1}$ &  $10^{-19}$\,erg\,s$^{-1}$\,cm$^{-2}$ & $\angstrom$           &  $10^{-19}$\,erg\,s$^{-1}$\,cm$^{-2}$ & $\angstrom$           \\ \hline \endhead

+53.15682-27.76716 & 7.982 & Deep/HST & $-18.71\pm0.05$ & $2.9\pm0.1$ & $7\pm1$ & $27\pm5$ & $7\pm1$ & $27\pm8$\\ 
+53.13347-27.76037 & 7.660 & Medium/HST & $-20.43\pm0.03$ & $21.9\pm0.6$ & $25\pm4$ & $13\pm2$ & $42\pm3$ & $22\pm3$\\ 
+53.16746-27.77201 & 7.276 & Deep/HST & $-17.00\pm0.25$ & $1.3\pm0.4$ & $15\pm7$ & $134\pm50$ & $23\pm2$ & $207\pm27$\\ 
+53.15579-27.81520 & 6.718 & Deep/HST & $-18.51\pm0.08$ & $2.9\pm0.4$ & $17\pm5$ & $77\pm22$ & $27\pm3$ & $122\pm27$\\ 
+53.16904-27.77884 & 6.633 & Deep/HST & $-18.76\pm0.04$ & $6.4\pm0.3$ & $21\pm3$ & $44\pm6$ & $22\pm5$ & $45\pm20$\\ 
+53.13492-27.77271 & 6.336 & Deep/HST & $-20.11\pm0.01$ & $25.3\pm0.8$ & $72\pm6$ & $39\pm3$ & $77\pm4$ & $41\pm4$\\ 
+53.17836-27.80098 & 6.327 & Medium/HST & $-18.75\pm0.18$ & $9.0\pm1.8$ & $57\pm20$ & $86\pm26$ & $43\pm12$ & $65\pm36$\\ 
+53.09517-27.76061 & 6.263 & Medium/HST & $-19.85\pm0.06$ & $26.4\pm1.7$ & $42\pm9$ & $22\pm5$ & $31\pm8$ & $16\pm8$\\ 
+53.08604-27.74760 & 6.204 & Medium/HST & $-18.93\pm0.14$ & $5.8\pm1.3$ & $29\pm13$ & $70\pm27$ & $48\pm21$ & $114\pm99$\\ 
+53.16062-27.77161 & 5.975 & Deep/HST & $-18.63\pm0.04$ & $6.7\pm0.3$ & $33\pm3$ & $72\pm6$ & $22\pm5$ & $47\pm20$\\ 
+53.12175-27.79763 & 5.937 & Deep/HST & $-19.83\pm0.01$ & $10.7\pm0.7$ & $123\pm13$ & $165\pm15$ & $106\pm9$ & $142\pm25$\\ 
+53.11041-27.80892 & 5.937 & Deep/HST & $-18.68\pm0.03$ & $9.6\pm0.5$ & $29\pm3$ & $44\pm5$ & $29\pm6$ & $43\pm19$\\ 
+53.16280-27.76084 & 5.917 & Medium/HST & $-19.60\pm0.08$ & $17.1\pm1.8$ & $90\pm17$ & $76\pm12$ & $112\pm19$ & $95\pm32$\\ 
+53.17655-27.77111 & 5.889 & Deep/HST & $-18.72\pm0.06$ & $11.2\pm1.1$ & $74\pm12$ & $95\pm12$ & $-$ & $-$\\ 
+53.11351-27.77284 & 5.814 & Deep/HST & $-18.18\pm0.05$ & $5.7\pm0.5$ & $60\pm8$ & $153\pm17$ & $38\pm5$ & $97\pm25$\\ 
+53.16713-27.79424 & 5.770 & Medium/HST & $-17.32\pm0.60$ & $3.8\pm2.2$ & $41\pm36$ & $159\pm106$ & $-$ & $-$\\ 
+53.17350-27.82507 & 5.609 & Medium/JWST & $-18.23\pm0.19$ & $2.8\pm1.2$ & $44\pm29$ & $238\pm120$ & $-$ & $-$\\ \hline 
+53.16477-27.77463 & 11.900 & Deep/HST & $-19.36\pm0.05$ & $1.6\pm0.2$ & $<5$ & $<23$ & $-$ & $-$\\ 
+53.15884-27.77349 & 9.870 & Deep/HST & $-18.45\pm0.11$ & $1.1\pm0.2$ & $<7$ & $<58$ & $-$ & $-$\\ 
+53.16735-27.80750 & 9.690 & Deep/HST & $-19.27\pm0.04$ & $3.1\pm0.1$ & $<5$ & $<14$ & $-$ & $-$\\ 
+53.11243-27.77461 & 9.440 & Deep/HST & $-20.36\pm0.01$ & $8.8\pm0.2$ & $<5$ & $<5$ & $-$ & $-$\\ 
+53.16446-27.80218 & 8.482 & Deep/HST & $-17.54\pm0.20$ & $0.5\pm0.1$ & $<8$ & $<161$ & $-$ & $-$\\ 
+53.08722-27.77706 & 7.580 & Medium/HST & $-19.03\pm0.20$ & $6.8\pm1.1$ & $<56$ & $<95$ & $-$ & $-$\\ 
+53.18343-27.79097 & 7.429 & Medium/JWST & $-20.00\pm0.05$ & $11.3\pm0.7$ & $<37$ & $<39$ & $-$ & $-$\\ 
+53.19105-27.79731 & 7.266 & Medium/JWST & $-19.15\pm0.10$ & $6.0\pm0.5$ & $<30$ & $<60$ & $-$ & $-$\\ 
+53.18714-27.80129 & 7.264 & Medium/JWST & $-18.41\pm0.20$ & $3.5\pm0.6$ & $<35$ & $<122$ & $-$ & $-$\\ 
+53.15283-27.80194 & 7.262 & Deep/HST & $-17.95\pm0.10$ & $2.0\pm0.3$ & $<8$ & $<48$ & $-$ & $-$\\ 
+53.18374-27.79390 & 7.262 & Medium/JWST & $-17.94\pm0.36$ & $2.9\pm0.8$ & $<40$ & $<170$ & $-$ & $-$\\ 
+53.16483-27.78826 & 7.250 & Medium/HST & $-19.08\pm0.18$ & $8.9\pm1.2$ & $<58$ & $<79$ & $-$ & $-$\\ 
+53.16172-27.78539 & 7.240 & Medium/HST & $-19.47\pm0.18$ & $7.7\pm1.0$ & $<65$ & $<103$ & $-$ & $-$\\ 
+53.16556-27.77266 & 7.240 & Medium/HST & $-18.32\pm0.29$ & $4.6\pm1.8$ & $<45$ & $<119$ & $-$ & $-$\\ 
+53.11833-27.76901 & 7.206 & Deep/HST & $-18.80\pm0.05$ & $5.1\pm0.3$ & $<10$ & $<25$ & $-$ & $-$\\ 
+53.13806-27.78186 & 7.140 & Medium/HST & $-18.66\pm0.22$ & $1.8\pm0.4$ & $<46$ & $<312$ & $-$ & $-$\\ 
+53.13423-27.76891 & 7.052 & Deep/HST & $-18.84\pm0.06$ & $7.6\pm0.5$ & $<12$ & $<19$ & $-$ & $-$\\ 
+53.17688-27.78156 & 7.002 & Medium/JWST & $-17.94\pm0.28$ & $1.2\pm0.5$ & $<31$ & $<334$ & $-$ & $-$\\ 
+53.18302-27.78946 & 6.951 & Medium/JWST & $-17.92\pm0.27$ & $1.4\pm0.5$ & $<30$ & $<271$ & $-$ & $-$\\ 
+53.11730-27.76408 & 6.930 & Deep/HST & $-18.69\pm0.07$ & $4.7\pm0.3$ & $<13$ & $<35$ & $-$ & $-$\\ 
+53.14771-27.71537 & 6.846 & Medium/HST & $-20.33\pm0.05$ & $22.3\pm1.3$ & $<45$ & $<26$ & $-$ & $-$\\ 
+53.11817-27.79302 & 6.800 & Medium/HST & $-19.01\pm0.13$ & $7.2\pm1.1$ & $<41$ & $<74$ & $-$ & $-$\\ 
+53.11634-27.76194 & 6.794 & Medium/HST & $-18.88\pm0.12$ & $7.3\pm1.1$ & $<34$ & $<61$ & $-$ & $-$\\ 
+53.15138-27.81917 & 6.711 & Deep/HST & $-17.96\pm0.07$ & $3.8\pm0.3$ & $<7$ & $<25$ & $-$ & $-$\\ 
+53.10538-27.72347 & 6.636 & Medium/HST & $-20.48\pm0.04$ & $27.0\pm1.4$ & $<63$ & $<31$ & $-$ & $-$\\ 
+53.15160-27.78791 & 6.629 & Medium/JWST & $-17.02\pm0.34$ & $1.2\pm0.5$ & $<20$ & $<208$ & $-$ & $-$\\ 
+53.16288-27.76928 & 6.624 & Deep/HST & $-17.61\pm0.20$ & $2.0\pm0.5$ & $<16$ & $<107$ & $-$ & $-$\\ 
+53.13743-27.76519 & 6.622 & Medium/HST & $-18.43\pm0.26$ & $6.8\pm2.0$ & $<51$ & $<97$ & $25\pm10$ & $48\pm40$\\ 
+53.16951-27.75331 & 6.620 & Medium/HST & $-19.34\pm0.11$ & $12.9\pm1.6$ & $<46$ & $<47$ & $-$ & $-$\\ 
+53.12731-27.78805 & 6.390 & Medium/HST & $-19.43\pm0.13$ & $18.4\pm2.1$ & $<62$ & $<46$ & $-$ & $-$\\ 
+53.12556-27.78676 & 6.390 & Medium/HST & $-18.36\pm0.34$ & $5.7\pm2.0$ & $<59$ & $<139$ & $-$ & $-$\\ 
+53.19660-27.81345 & 6.340 & Medium/HST & $-18.24\pm0.28$ & $5.4\pm1.0$ & $<42$ & $<105$ & $-$ & $-$\\ 
+53.17582-27.77446 & 6.336 & Deep/HST & $-18.74\pm0.06$ & $6.4\pm0.6$ & $<15$ & $<32$ & $-$ & $-$\\ 
+53.16660-27.77240 & 6.330 & Deep/HST & $-18.56\pm0.05$ & $6.7\pm0.4$ & $<9$ & $<18$ & $-$ & $-$\\ 
+53.15516-27.76072 & 6.314 & Medium/HST & $-19.42\pm0.05$ & $15.9\pm0.8$ & $<27$ & $<23$ & $-$ & $-$\\ 
+53.16613-27.77204 & 6.300 & Medium/HST & $-18.47\pm0.25$ & $7.3\pm1.9$ & $<52$ & $<98$ & $-$ & $-$\\ 
+53.16238-27.80332 & 6.298 & Deep/HST & $-16.33\pm0.62$ & $1.0\pm0.6$ & $<15$ & $<195$ & $-$ & $-$\\ 
+53.08311-27.78635 & 6.260 & Medium/HST & $-19.72\pm0.09$ & $17.8\pm2.2$ & $<62$ & $<48$ & $-$ & $-$\\ 
+53.16902-27.80079 & 6.250 & Medium/HST & $-18.95\pm0.16$ & $10.9\pm1.5$ & $<48$ & $<61$ & $-$ & $-$\\ 
+53.20800-27.79005 & 6.180 & Medium/HST & $-19.25\pm0.16$ & $5.2\pm1.8$ & $<90$ & $<242$ & $-$ & $-$\\ 
+53.15613-27.77584 & 6.107 & Deep/HST & $-19.02\pm0.04$ & $8.4\pm0.6$ & $<11$ & $<19$ & $-$ & $-$\\ 
+53.15953-27.77152 & 6.100 & Medium/HST & $-20.43\pm0.04$ & $40.7\pm1.7$ & $<43$ & $<15$ & $-$ & $-$\\ 
+53.19588-27.76843 & 6.060 & Medium/HST & $-18.80\pm0.16$ & $7.7\pm1.6$ & $<41$ & $<75$ & $-$ & $-$\\ 
+53.17324-27.79567 & 6.000 & Medium/HST & $-19.66\pm0.06$ & $15.3\pm1.1$ & $<33$ & $<31$ & $-$ & $-$\\ 
+53.17264-27.76706 & 5.990 & Medium/HST & $-19.02\pm0.12$ & $9.5\pm1.4$ & $<42$ & $<63$ & $-$ & $-$\\ 
+53.19938-27.79627 & 5.980 & Medium/HST & $-19.44\pm0.09$ & $17.2\pm1.4$ & $<42$ & $<35$ & $-$ & $-$\\ 
+53.14902-27.78070 & 5.956 & Medium/JWST & $-17.85\pm0.34$ & $4.4\pm1.6$ & $<37$ & $<121$ & $-$ & $-$\\ 
+53.11911-27.76080 & 5.949 & Deep/HST & $-19.42\pm0.02$ & $17.7\pm0.5$ & $<7$ & $<6$ & $-$ & $-$\\ 
+53.11264-27.77262 & 5.934 & Medium/HST & $-18.41\pm0.22$ & $4.9\pm1.7$ & $<41$ & $<120$ & $-$ & $-$\\ 
+53.12654-27.81809 & 5.932 & Deep/HST & $-18.93\pm0.04$ & $10.7\pm0.6$ & $<10$ & $<13$ & $-$ & $-$\\ 
+53.13044-27.80236 & 5.930 & Medium/HST & $-19.17\pm0.11$ & $7.8\pm1.8$ & $<71$ & $<131$ & $-$ & $-$\\ 
+53.15218-27.77840 & 5.927 & Medium/JWST & $-16.99\pm0.43$ & $0.9\pm0.5$ & $<19$ & $<316$ & $-$ & $-$\\ 
+53.10547-27.76115 & 5.925 & Medium/HST & $-20.09\pm0.05$ & $18.0\pm1.4$ & $<42$ & $<33$ & $-$ & $-$\\ 
+53.12259-27.76057 & 5.920 & Deep/HST & $-19.64\pm0.02$ & $24.0\pm0.6$ & $<9$ & $<6$ & $-$ & $-$\\ 
+53.14987-27.75283 & 5.919 & Medium/HST & $-19.39\pm0.07$ & $13.7\pm1.0$ & $<33$ & $<35$ & $-$ & $-$\\ 
+53.17752-27.80252 & 5.860 & Medium/HST & $-19.15\pm0.11$ & $15.2\pm1.3$ & $<40$ & $<38$ & $-$ & $-$\\ 
+53.14197-27.75523 & 5.827 & Medium/HST & $-19.70\pm0.08$ & $24.8\pm1.6$ & $<41$ & $<24$ & $-$ & $-$\\ 
+53.16730-27.80287 & 5.818 & Deep/HST & $-17.89\pm0.09$ & $3.2\pm0.3$ & $<9$ & $<43$ & $-$ & $-$\\ 
+53.15407-27.76607 & 5.807 & Deep/HST & $-18.94\pm0.04$ & $10.3\pm0.6$ & $<9$ & $<13$ & $-$ & $-$\\ 
+53.12554-27.75505 & 5.780 & Medium/HST & $-19.35\pm0.11$ & $9.3\pm1.7$ & $<57$ & $<90$ & $-$ & $-$\\ 
+53.13385-27.77858 & 5.778 & Medium/JWST & $-16.72\pm0.73$ & $2.1\pm1.3$ & $<28$ & $<200$ & $-$ & $-$\\ 
+53.14505-27.81643 & 5.774 & Deep/HST & $-18.05\pm0.10$ & $4.7\pm0.7$ & $<13$ & $<40$ & $-$ & $-$\\ 
+53.11537-27.81477 & 5.770 & Deep/HST & $-19.32\pm0.03$ & $14.4\pm0.7$ & $<12$ & $<13$ & $-$ & $-$\\ 
+53.11775-27.81653 & 5.761 & Medium/JWST & $-17.84\pm0.32$ & $6.2\pm1.7$ & $<36$ & $<85$ & $-$ & $-$\\ 
+53.21160-27.79639 & 5.760 & Medium/HST & $-19.35\pm0.10$ & $7.3\pm1.1$ & $<43$ & $<88$ & $-$ & $-$\\ 
+53.13059-27.80771 & 5.617 & Deep/HST & $-18.58\pm0.09$ & $9.9\pm1.3$ & $<18$ & $<27$ & $-$ & $-$\\ 
+53.18064-27.82239 & 5.606 & Medium/JWST & $-20.12\pm0.03$ & $29.3\pm1.4$ & $<29$ & $<15$ & $-$ & $-$\\ 
\end{longtable}